\date{\today}

\documentclass[aps,prb,twocolumn,superscriptaddress,groupedaddress]{revtex4}
\usepackage{dcolumn,graphicx,amsmath,amssymb,amsfonts,algorithm,algpseudocode}
\usepackage[colorlinks=true,citecolor=blue,urlcolor=blue,linkcolor=blue]{hyperref}

\begin{document}

\title{
Simple Computation of the MultiExp Gaussian Quadrature in Double Precision
}

\author{Robert M. Parrish}
\email{rob.parrish@qcware.com}
\affiliation{
QC Ware Corporation, 195 Page Mill Rd, Ste.~113, Palo Alto, CA 94306, USA
}

\begin{abstract}
The MultiExp Gaussian quadrature rule proposed by Gill and Chien has many
compelling properties for the integration of radial integrands encountered in
electronic structure, but thus far has only been computed up to $N = 50$, with
some nontrivial holes in the existing tables for $27 \leq N < 50$.  In this
work, a simple recipe is developed to compute the MultiExp Gaussian quadrature
for larger $N$ at close to the double precision machine epsilon.  Notably, this
recipe uses only double precision operations, so no computer algebra systems or
arbitrary precision libraries are needed. There are three primary outcomes: (A)
The MultiExp tables for all $N \leq 1000$ are presented in a supplement to this
report - these might prove useful in building better medium and high quality
density functional theory quadrature grids (B) these tables are compared to the
known Gill and Chien tables with excellent agreement obtained modulo a
noticeable problem with the weights of the largest $N = 50$ Gill and Chien table
and (C) it may be the case that the Boley-Golub plus extremely high-order
Gauss-Legendre quadrature approximation recipe developed in this work could be
used to automatically derive other quadrature rules throughout electronic
structure theory and beyond.
\end{abstract}

\maketitle

\section{Introduction}

The efficient quadrature-based integration of complicated integrands in the
radial coordinate of the form $I \equiv \int_{0}^{\infty} \mathrm{d} r \ r^2
f(r) \approx \sum_{i} u_{i} f(r_i)$ is a crucial task in molecular electronic
structure theory. As just one example, this is a key step in the
``molecule-shaped'' integration of the nonlinear Kohn-Sham exchange-correlation
potential over of density functional theory. Moreover, this step seems to have
more heuristic choice involved than the other steps. Molecular quadratures of
the Becke polyatomic grid type\cite{becke1988multicenter} can usually be thought
of in terms of three steps: (1) atomic partitioning, (2) solid angle
integration, and (3) radial integration. Despite a fair amount of high-quality
work on (1), the widely prevailing sentiment is that Becke's original
choice\cite{becke1988multicenter} of a thrice-iterated smooth cubic cutoff
function in confocal elliptic coordinates provides a heuristically optimal
smooth Voronoi tessellation of the molecule for this consideration (note that a
small ``bump-function'' modification of this procedure by Stratmann, Scuseria,
and Frisch\cite{stratmann1996achieving} achieves essentially the same quality of
partition function, albeit with the possibility of strict linear scaling
computation). (2) involves integration of medium-order spherical harmonics over
the solid angle, with a strong desire to preserve at least octahedral symmetry,
leading to Lebedev-Laikov
grids\cite{lebedev1976quadratures,lebedev1999quadrature} as the widely
prevailing choice.

\begin{figure}[h!]
\begin{center}
\includegraphics[width=3.4in]{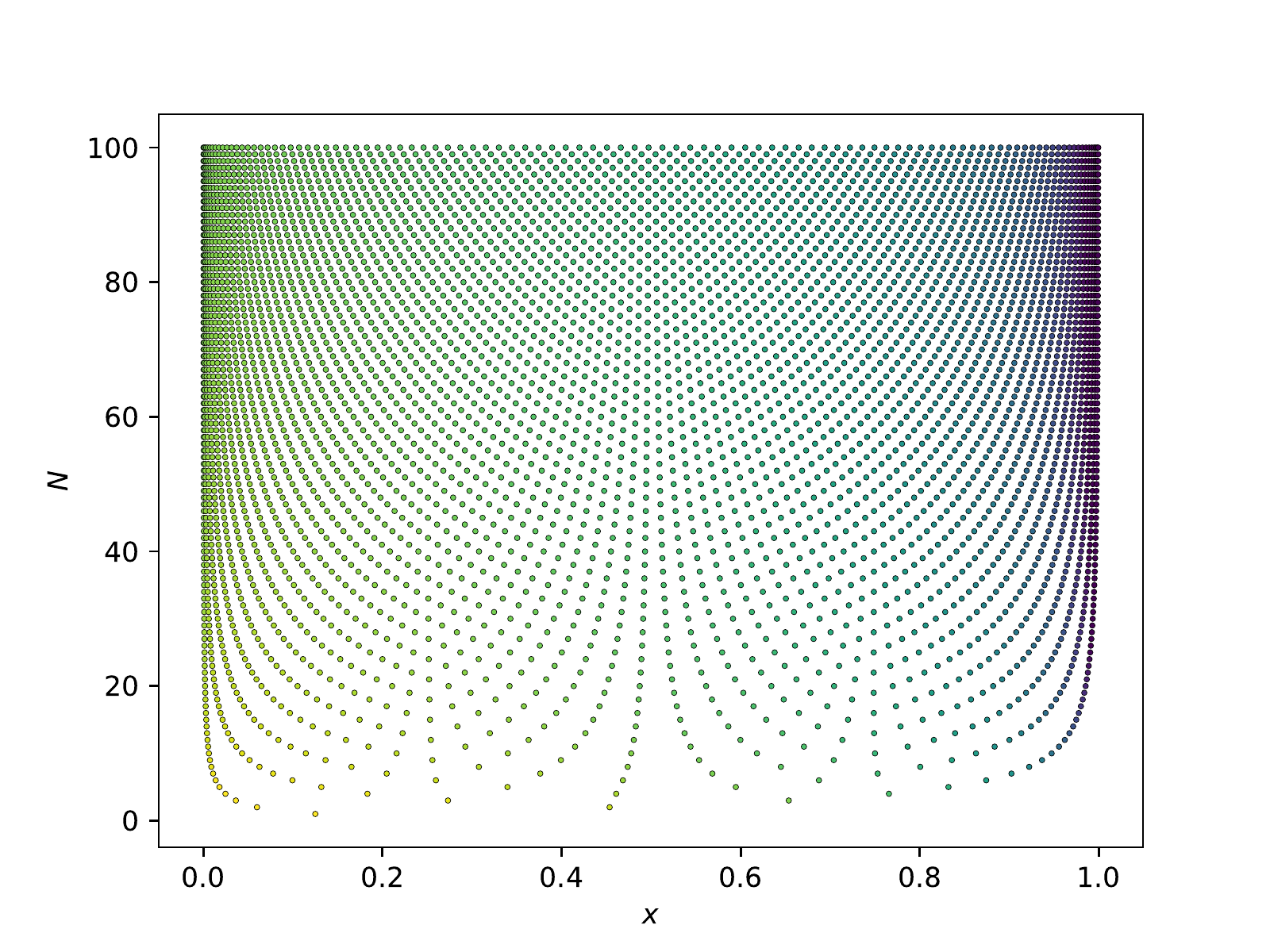}
\caption{Visual representation of the finished log-squared quadrature grids for $1
\leq N \leq 100$ obtained in this work. The colors represent the logarithms of
the weights, with $10^{0}$ appearing as yellow and $10^{-6}$ appearing as
purple. Some weights exceed this range and are clamped to the endpoint colors.}
\label{fig:multiexp}
\end{center}
\end{figure}

For (3), many proposals have been made: ``classic'' Gauss-Laguerre,
Becke,\cite{becke1988multicenter} Handy,\cite{murray1993quadrature}
Ahlrichs,\cite{treutler1995efficient} Knowles,\cite{mura1996improved} Mitani
and Yoshioka's double exponential,\cite{mitani2012numerical} and Shizgal et
al's Gauss-Maxwell,\cite{shizgal2017computation} among others, and all of these
seem to have their own pros and cons, e.g., there is seemingly no clear choice
of a universal radial quadrature rule. However, there is one more radial
quadrature rule that is highly compelling: the multi-exponential or
``MultiExp'' quadrature of Gill and Chien.\cite{gill2003radial} The MultiExp
quadrature starts from the ``log-squared''  Gaussian quadrature rule for the
unique $N$-node grid of points and weights $\{ < x_i, w_i > \}$ that exactly
integrate polynomials of order $2N - 1$ over $x \in [0, 1]$ when multiplied by
the weight function $\ln^2 (x)$, i.e.,
\begin{equation}
\int_{0}^{1}
\mathrm{d} x \
\ln^2(x)
f (x)
=
\sum_{i=1}^{N}
w_{i}
f(x_i)
\
\forall
\
f (x)
\in
\mathcal{P}_{2N - 1} (x)
\end{equation}
The log-squared quadrature rules for $N \leq 100$ are depicted in Fig.
\ref{fig:multiexp} - the precise determination of the log-squared quadratures up
to $N = 1000$ is the primary outcome of the present work.

With an ``atomic size'' parameter $R$, the transformation,
\begin{equation}
r_{i}
\equiv
-
R
\ln (x_{i})
\end{equation}
and,
\begin{equation}
u_{i}
\equiv
R^3
\frac{w_{i}}{x_{i}} 
\end{equation}
yields the desired radial MultiExp quadrature rule,
\begin{equation}
\int_{0}^{\infty}
\mathrm{d} r \
r^2
f (r)
\approx
\sum_{i=1}^{N}
u_{i}
f (r_i)
\end{equation}
Just as the log-squared quadrature is exact for polynomials in $x$ integrands
multiplied by $\ln^2 (x)$, the MultiExp radial quadrature grid is exact for
integrands in $r$ multiplied by $r^2$ of the ``multi-exponential'' form,
\begin{equation}
f (r)
= 
\sum_{k = 1}^{2 N}
a_{k}
\exp(-k r / R)
\
\forall
\
\{
a_{k}
\}
\in
\mathbb{R}^{2N}
\end{equation}
This multi-exponential form is heuristically well-suited to radial integrands
involving the electronic density that are commonly encountered in electronic
structure theory. In practice, the MultiExp quadratures perform remarkably well
for density-involving integrands, e.g., forming the basis for the
``good-for-the-money'' SG-0 grid\cite{chien2006sg} that is widely used in
electronic structure.

One major barrier to further adoption of the MultiExp quadrature is that the
accurate computation of the quadrature rules appears to be exceedingly difficult
with standard techniques. Gill has tabulated the rules to 15 absolute decimal
digits for $N \in \{1, \ldots, 26, 30, 35, 40, 45, 50\}$ on his research group's
website,\cite{gill_chien} and these specific rules are widely available in open-source and
proprietary electronic structure codes. In the Gill and Chien MultiExp paper,\cite{gill2003radial}
the recipe to determine these weights is stated as ``[B]y inverting the Cholesky
triangle of the associated Gram matrix, we have constructed the $Q_{n} (x)$ [the
orthogonal polynomials for $\ln^2 (x)$] for $1 \leq n \leq 27$. The associated
roots and weights \ldots{} are listed in Table 1.'' This together with the
integral example definition of $Q_{0} (x) = 1$, $Q_{1} (x) = 8x - 1$, $Q_{2} (x)
= 7992 x^2 - 4104 x + 217$ (neither monic nor normalized) provided in a footnote
suggest that an integer coefficient computer algebra system for polynomials was
used to derive the explicit form of the orthogonal polynomials. The roots of
these polynomials and weights (related to the derivative of the polynomial at
each root) were then likely extracted by an arbitrary-precision numerical
procedure to obtain the tabulated quadratures to near double-precision accuracy.
Note that exact integer representation of a family of orthogonal polynomials is
often considered to be a last resort, as the required RAM and compute time to
store and manipulate the integer coefficients may grow exponentially with order.
It is quite impressive that this approach has already made it out to $N=45$ or
$N=50$.  (However, the present work indicates there may be a non-trivial problem
with the grid Gill and Chien report for $N=50$, see below).

In this work we will (1) describe a simple recipe to compute the MultiExp
quadrature rules to close to the double precision machine epsilon using only
double precision computation and (2) present the MultiExp roots and weights for
$N \leq 1000$. Note that below, we prefer to define the log-squared Gaussian
quadrature as \emph{the} MultiExp quadrature - the context of integrating in $x$
or $r$ will resolve this nuance in the notation.

\section{Theory}

\subsection{The Boley-Golub Algorithm}

The starting point for our method is the Boley-Golub
algorithm\cite{boley1987survey} to transform a quadrature rule (Gaussian or
not) in one measure to a true Gaussian quadrature rule in a different measure.
An additional explanatory reference on the Boley-Golub algorithm is as
follows.\cite{fernandes2006gaussian}

We seek the $N$-point Gaussian quadrature rule $\{ <x_i, w_i> \}$ for
integration over $x$ from $a$ to $b$ with weight function $\mu (x)$,
\begin{equation}
\int_{a}^{b}
\mathrm{d} x \
\mu (x)
f (x)
=
\sum_{i=1}^{N}
w_{i}
f (x_i)
:
\
f (x)
\in
\mathcal{P}_{2N - 1} (x)
\end{equation}

Let us say we have another set of $Q$-point quadrature rules $\{ < \xi_{i}, q_{i}
> \}$ for the measure integral,\footnote{Here there is some subtlety. We mean
that in the high-$Q$ limit, we can replace the continuous measure with its
discrete counterpart,
\begin{equation}
\int_{a}^{b}
\mathrm{d} x \
\mu(x) 
= 
\lim_{Q \rightarrow \infty}
\sum_{i=1}^{Q} 
q_{i} \mu (\xi_{i}) \delta(x - \xi_{i})
\end{equation}
And moreover that we can do this for the various needed moments involving this
measure.
}
\begin{equation}
\int_{a}^{b}
\mathrm{d} x \
\mu (x)
\approx
\sum_{i=1}^{Q}
q_{i}
\mu (\xi_{i})
\end{equation}

Now we can construct the $Q+1$-dimensional square symmetric matrix,
\begin{equation}
\hat W
\equiv
\left [
\begin{array}{r|rrrr}
1 & \sqrt{\omega_{0}} & \sqrt{\omega_{1}} & \sqrt{\omega_{2}} & \hdots \\
\hline
\sqrt{\omega_{0}} & \xi_{0} & 0 & 0 & \hdots \\
\sqrt{\omega_{1}} & 0 & \xi_{1} & 0 & \hdots \\
\sqrt{\omega_{2}} & 0 & 0 & \xi_{2} & \hdots \\
\vdots & \vdots & \vdots & \vdots & \ddots \\
\end{array}
\right ]
:
\
\sqrt{
\omega_{i}
}
\equiv
\sqrt{
q_{i}
\mu (\xi_i)
}
\end{equation}

This orthogonally tridiagonalizes in the large $Q$ limit via,
\begin{equation}
\hat T
=
\hat Q^{\dagger}
\hat T
\hat Q
:
\
\hat Q
\hat Q^{\dagger}
=
\hat I
\end{equation}
to,
\begin{equation}
\hat T
\equiv
\left [
\begin{array}{r|rrrr}
1 & \sqrt{b_{0}} & 0 & 0 & \hdots \\
\hline
\sqrt{b_{0}} & d_{0} & \sqrt{b_{1}} & 0 & \hdots \\
0 & \sqrt{b_{1}} & d_{1} & \sqrt{b_{2}} & \hdots \\
0 & 0 & \sqrt{b_{2}} & d_{2} & \hdots \\
\vdots & \vdots & \vdots & \vdots & \ddots \\
\end{array}
\right ]
\equiv
\left [
\begin{array}{r|r}
1 & \vec c_{0}^{\dagger} \\
\hline
\vec c_{0} & \hat J \\
\end{array}
\right ]
\end{equation}
Note that for our purposes, the phases of the tridiagonalization can be
canonicalized by requiring that the superdiagonal elements all be positive.

The second full minor $\hat J$ is the Golub-Welsch Jacobian matrix. The entry
$T_{1,0}$ is $\sqrt{b_{0}}$, the square root of the zeroth moment,
\begin{equation}
b_{0}
=
\int_{a}^{b}
\mathrm{d} x \
\mu (x)
\end{equation}

\subsection{The Golub-Welsch Algorithm}

As soon as $N+1$ rows and columns of $\hat T$ are uncoupled in the
tridiagonalization procedure, the symmetric, tridiagonal Golub-Welsch Jacobian
matrix $\hat J_{(N)}$ is available, i.e., full tridiagonalization of $\hat T$ is
not required for $N < Q$.

At this point, we turn to the famous Golub-Welsch
algorithm,\cite{golub1969calculation} which is succinctly stated as follows:
Diagonalize the symmetric, tridiagonal Golub-Welsch Jacobian matrix,
\begin{equation}
\hat J_{(N)} 
=
\hat U_{(N)}
\hat x^{(N)}
\hat U_{(N)}^{\dagger}
\end{equation}
and obtain the $N$-node Gaussian quadrature roots as the eigenvalues of $\hat
J_{(N)}$, i.e., $\{ x_i \} = \hat x^{(N)}$. The corresponding Gaussian
quadrature weights are obtained from the first row of the eigenvectors as,
\begin{equation}
w_{i}
=
b_{0}
\left (
\hat U_{(N)} [0, i]
\right )^2
\forall
\
i
\in
1 \ldots N
\end{equation}

\subsection{Accuracy Considerations}

\textbf{Stability:} The orthogonal tridiagonalization procedure of the Boley-Golub
algorithm and the diagonalization of the Golub-Welsch algorithm both involve
orthogonal transforms that are typically found to be stable and precise to a few
ulp. Specifically, we use a standard system LAPACK DSYEV call for the needed
eigenvalues and eigenvectors in Golub-Welsch (it is relatively easy to reduce the
Golub-Welsch diagonalization computational effort by exploiting the fact that
$\hat J_{(N)}$ is already tridiagonal to achieve linear scaling in this step,
but this is overkill for the $N \leq 1000$ we will be considering within this
work). For the partial tridiagonalization step of the Boley-Golub algorithm, we
use a handwritten algorithm using Householder reflectors, with the only unusual
concession to numerical stability being the use of Kahan-Babushka-Klein
summation\cite{kahan1965pracniques,klein2006generalized} in the very long dot
products of size $Q$. With these considerations, Boley-Golub + Golub-Welsch can
be considered to be precise up to a few ulp in double precision.

\textbf{Accuracy:} With stability out of the way, we now consider the
primary numerical consideration of the full procedure: accuracy. The only formal
approximation made in the whole procedure above is that the $Q$-point quadrature
rule $\{ < \xi_{i}, q_{i} > \}$ used in Boley-Golub is an approximate quadrature
rule for the target metric integral. In formal practice, the accuracy of the
overall $N$-node procedure is governed by the accuracy of resolving the set of
$2 N$ moments,
\begin{equation}
M^{(k)}
\equiv
\int_{a}^{b}
\mathrm{d} x \
\mu (x)
x^{k}
\approx
\sum_{i=1}^{Q}
q_{i}
\mu (\xi_i)
\xi_i^{k}
\
\forall
\
k \in [0, 2N - 1]
\end{equation}
The reason for this is that the Golub-Welsch Jacobi matrix is formally
completely defined by the lowest $2N$ moments $M^{(k)}$. However, we typically
eschew explicit manipulations of high-order moments in numerical practice due to
atrocious stability considerations, e.g., we use a Boley-Golub-construction of
the Golub-Welsch Jacobian rather than a moment-based construction. In our
specific prescription below, we will find that $M^{(0)} = b_{0}$ converges
dramatically more slowly with $Q$ than the other moments, and thus the accuracy
in $M^{(0)}$ can be used as a rough yardstick for the quality of our $Q$-point
Boley-Golub approximation that is seemingly independent of $N$.

\textbf{Accuracy Special Case:} One special case is worth talking about: A
$Q \geq N$-point Gaussian quadrature rule for the weight function $\mu (x)$
definitionally exactly integrates all of the lowest $2 Q \geq 2 N$ moments, and
therefore will yield an exact Boley-Golub + Golub-Welsch procedure (note that
there is a minor detail that the weight function is already built into the
Gaussian quadrature weights, so we must substitute $\sqrt{\omega_{i}} \equiv
\sqrt{w_{i}}$ for this case). This seems at first glance to be begging the
question - we are saying that a Gaussian quadrature is a great way to discover
the same or smaller Gaussian quadratures. But it turns out to be very useful to
help validate a proposed grid: running Boley-Golub + Golub-Welsch from a
proposed highly-accurate $Q$-point Gaussian quadrature rule should reproduce all
$N \leq Q$ Gaussian quadrature rules to nearly the machine epsilon, and any
deviation from this behavior is indicative of a serious problem with the
proposed $Q$-point Gaussian quadrature.

\subsection{High-Order Gauss-Legendre Quadrature}

To close the procedure and compute the MultiExp quadrature, all that remains is
to find an accurate arbitrary $Q$-node quadrature rule $\{ < \xi_i, q_i > \}$ for
the specific MultiExp moments,
\begin{equation}
M^{(k)}
\equiv
\int_{0}^{1}
\mathrm{d} x \
\ln^2 (x)
x^{k}
\approx
\sum_{i=1}^{Q}
q_{i}
\ln^2 (\xi_i)
\xi_{i}^{k}
=
\frac{2}{(k + 1)^3}
\end{equation}
These are a numerical bear: $k = 0$ is plagued by the log-squared pole at $x =
0$, while for $k > 0$ the pole is extinguished and increasingly higher effort is
required for the transients that gradually emerge in the middle of the
integration domain.  There are seemingly very few general-purpose quadrature
rules that can handle this behavior. Indeed, this may be the reason MultiExp is
so compelling - it is \emph{the} quadrature rule that handles this behavior.

After many failed attempts, we have found that extremely high-order
Gauss-Legendre quadrature rules can crack the problem. The ``classic'' $Q$-node
Gauss-Legendre quadrature rule is $\{ < x_i , w_i > \}$ where,
\begin{equation}
\int_{-1}^{+1}
\mathrm{d} x \
f (x)
=
\sum_{i=1}^{Q}
w_{i}
f (x_i)
\
\forall
\
f (x)
\in 
\mathcal{P}_{2 Q - 1} (x)
\end{equation}
We can then substitute this into our Boley-Golub + Golub-Welsch procedure via,
\begin{equation}
\xi_{i} 
\equiv
\frac{1}{2}
[x_{i} + 1]
,
\
q_{i}
\equiv
\frac{1}{2}
w_{i}
\end{equation}
\begin{equation}
\sqrt{
\omega_{i}
}
\equiv
\sqrt{q_{i}}
| \ln (\xi_i) |
\end{equation}
For reasons discussed momentarily, we actually prefer to substitute a nonlinear
remapping of the interval, for some paramter $\gamma$,
\begin{equation}
\xi_{i}'
\equiv
\xi_{i}^{\gamma}
,
\
q_{i}'
\equiv
\gamma
\xi_{i}^{\gamma - 1}
q_{i}
\end{equation}
With $\gamma = 1$ (no nonlinear remapping), we find that the moments converge
slowly but regularly, with $M^{(0)}$ dramatically dominating the convergence
behavior presumably due to the log-squared pole at $x = 0$ for $k = 0$. We have
managed to converge the MultiExp quadrature for $N \leq 100$ to almost the
double machine precision with $\gamma = 1$, but this requires an extremely high
cost of $Q \sim 2\times 10^{8}$. Using $\gamma = 2$ quadratically clusters more
of the Gauss-Legendre points near the difficult $x = 0$ portion of the interval
without significantly compromising the sampling near $x = 1$. We find that this
dramatically increases the convergence rate of the lower moments without
compromising the stability of the overall approach, e.g., allowing for double
precision convergence to be obtained by $Q \sim 10^{5}$.  Unless otherwise
specified, all results below use $\gamma = 2$.

The last step needed to close the procedure is to obtain a recipe to efficiently
obtain the $Q$-node Gauss-Legendre quadrature rule $\{ < x_i, w_i > \}$ to very
close to the double machine precision. Thanks to the excellent 2014 work of
Bogaert,\cite{bogaert2014iteration} this is a solved problem. Bogaert's paper
provides an explicit, stable, non-iterative double-precision recipe to compute
the Gauss-Legendre quadrature nodes and weights to the double machine precision
using only standard scalar math library functions. This procedure can be
implemented in an afternoon using $<200$ lines of C++ using only the standard
library. An entertaining ``race report'' on the highly interesting topic of
high-order Gauss-Legendre quadratures is provided by
Townsend.\cite{townsend2015race}

\section{Implementation}

A simple C++ implementation of the above procedure was developed in the
proprietary QC Ware \texttt{Tachyon} electronic structure library. Bogaert's
fast asymptotic Gauss-Legendre quadrature method was implemented from scratch in
a way that essentially mirrors Bogaerts paper, albeit that we find Chebyshev
conditioning of the $F$ and $W$ intermediates to not be required if the
lower-half of the Gauss-Legendre nodes are explicitly computed and then the
upper-half of the Gauss-Legendre nodes (which are less accurate) are obtained by
symmetry. As with Bogaert, we believe that the Gauss-Legendre quadrature rules
computed with this rule are accurate to a few ulp. A partial Householder
tridiagonalization subroutine was developed to compute $T_{(N)}$ from $\hat W$
without prohibitive explicit storage of $W$. Our rather naive implementation of
this method uses $\mathcal{O}(2NQ)$ words of RAM, and uses Kahan-Babushka-Klein
summation for the extremely long dot products over $Q$ encountered in this work.
The resulting $T_{(N)}$ matrices are stored for later use by Golub-Welsch
(the highest $T_{(N)}$ can be stored and subsets used for all $N' \leq N$
MultiExp tables desired). The Golub-Welsch code is a vanilla dense LAPACK SYEV
eigensolve.  

\section{Results}

\subsection{Convergence of Moments}

\begin{figure}[h!]
\begin{center}
\includegraphics[width=3.4in]{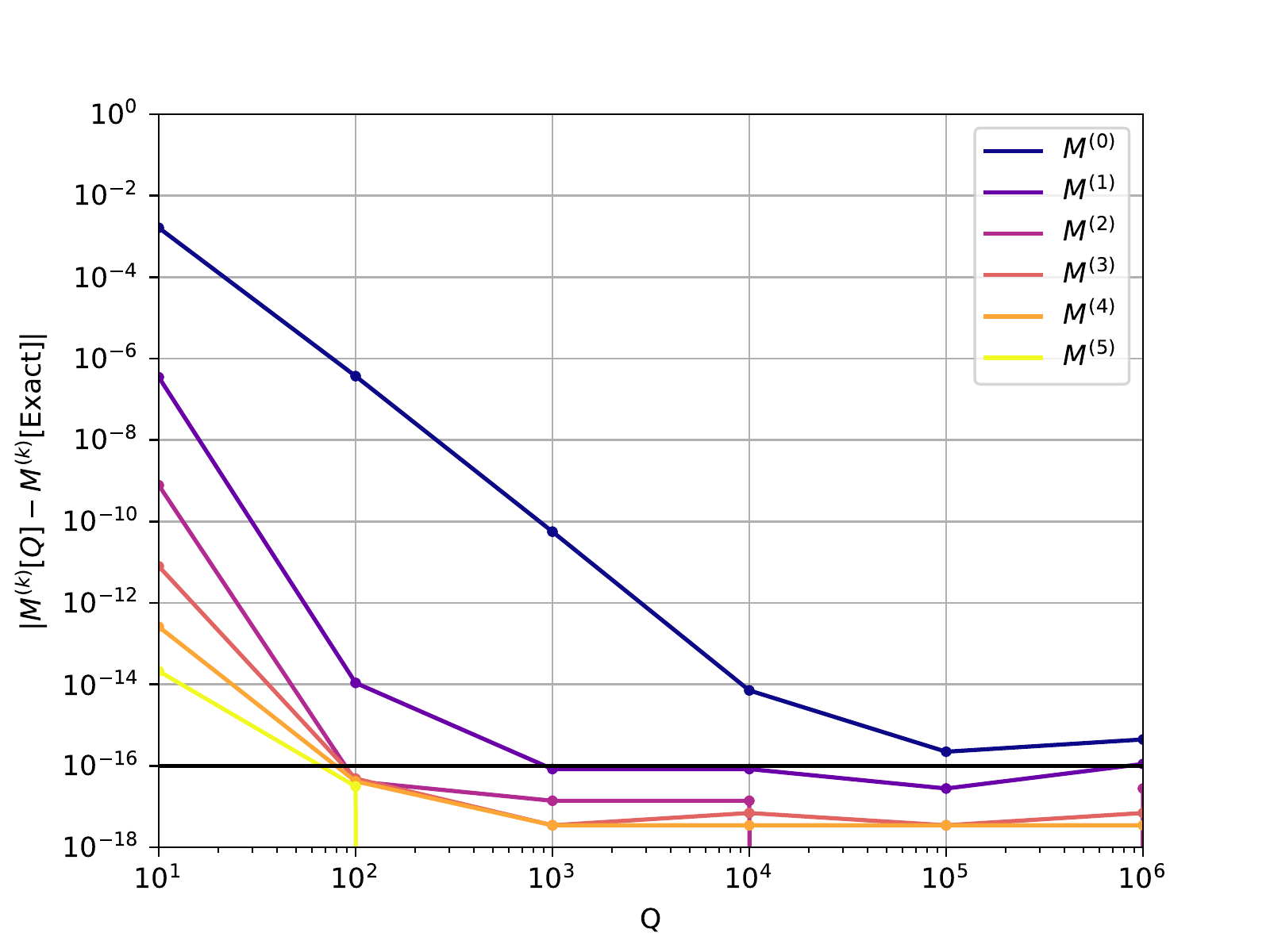}
\caption{Convergence of the MultiExp moments $M^{(k)}$ with respect to the size
of the Gauss-Legendre grid $Q$. The exact moments are $M^{(k)} \equiv
\int_{0}^{1} \mathrm{d} x \ \ln^2 (x) x^{k} = 2 / (k + 1)^3$.}
\label{fig:moments}
\end{center}
\end{figure}

Figure \ref{fig:moments} shows the convergence of the first few MultiExp moments
$M^{(k)}$ with respect to the size of the Gauss-Legendre grid $Q$ used in our
Boley-Golub + Golub-Welsch procedure. The convergence is clearly dominated by
$M^{(0)}$, which appears to converge as roughly $\mathcal{O} (Q^{-4})$ with our
selected $\gamma = 2$. Full double precision convergence is clearly obtained by
$Q = 10^{5}$.

\subsection{Convergence of Grids}

\begin{figure}[h!]
\begin{center}
\includegraphics[width=3.4in]{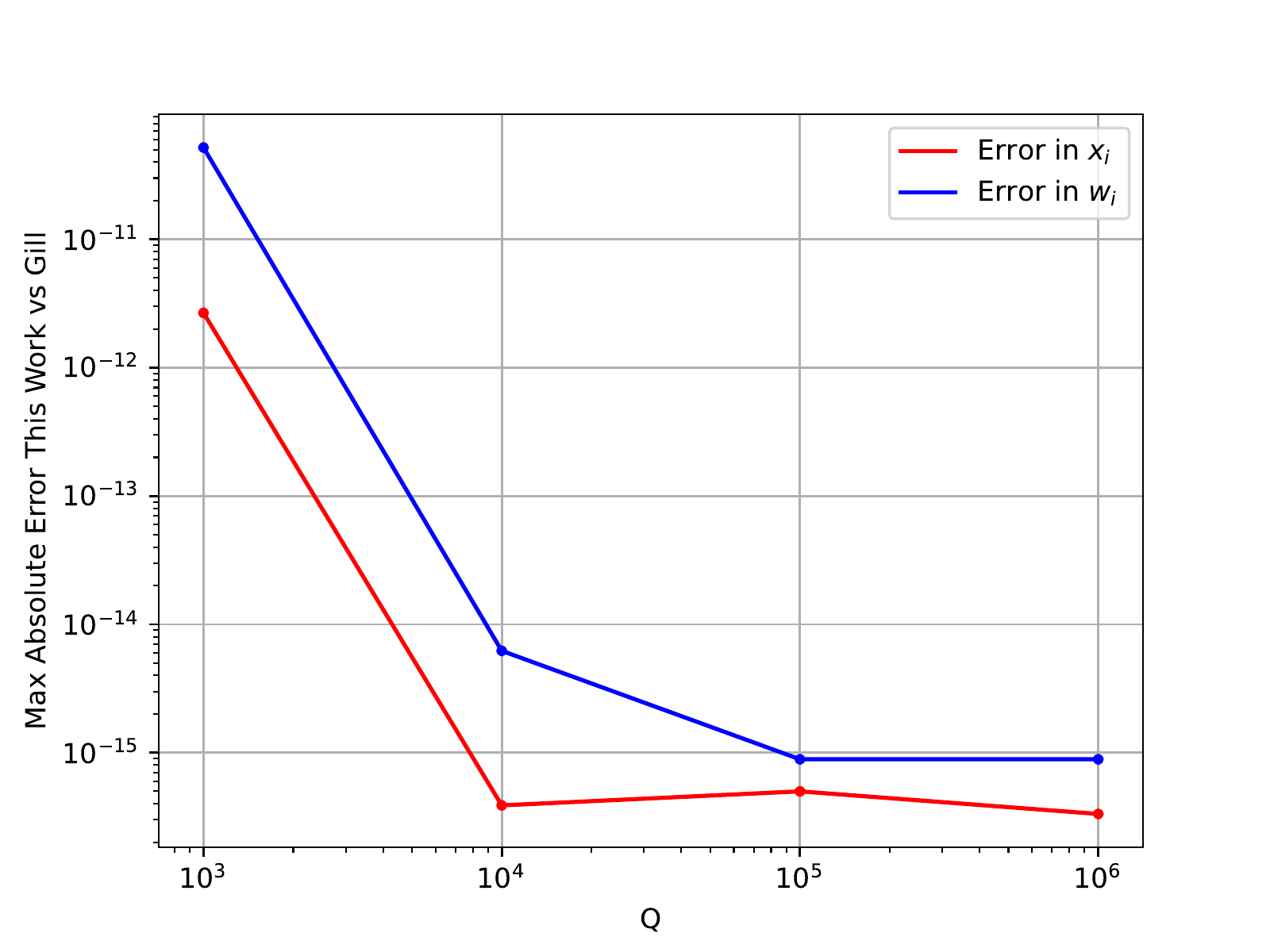}
\caption{Convergence of the MultiExp grid roots and weights with respect to the
size of the Gauss-Legendre grid $Q$, for $N = 2$. The reference is the tabulated
grid from Gill and Chien. The error characteristic is the maximum absolute
deviation.}
\label{fig:delta-2}
\end{center}
\end{figure}

\begin{figure}[h!]
\begin{center}
\includegraphics[width=3.4in]{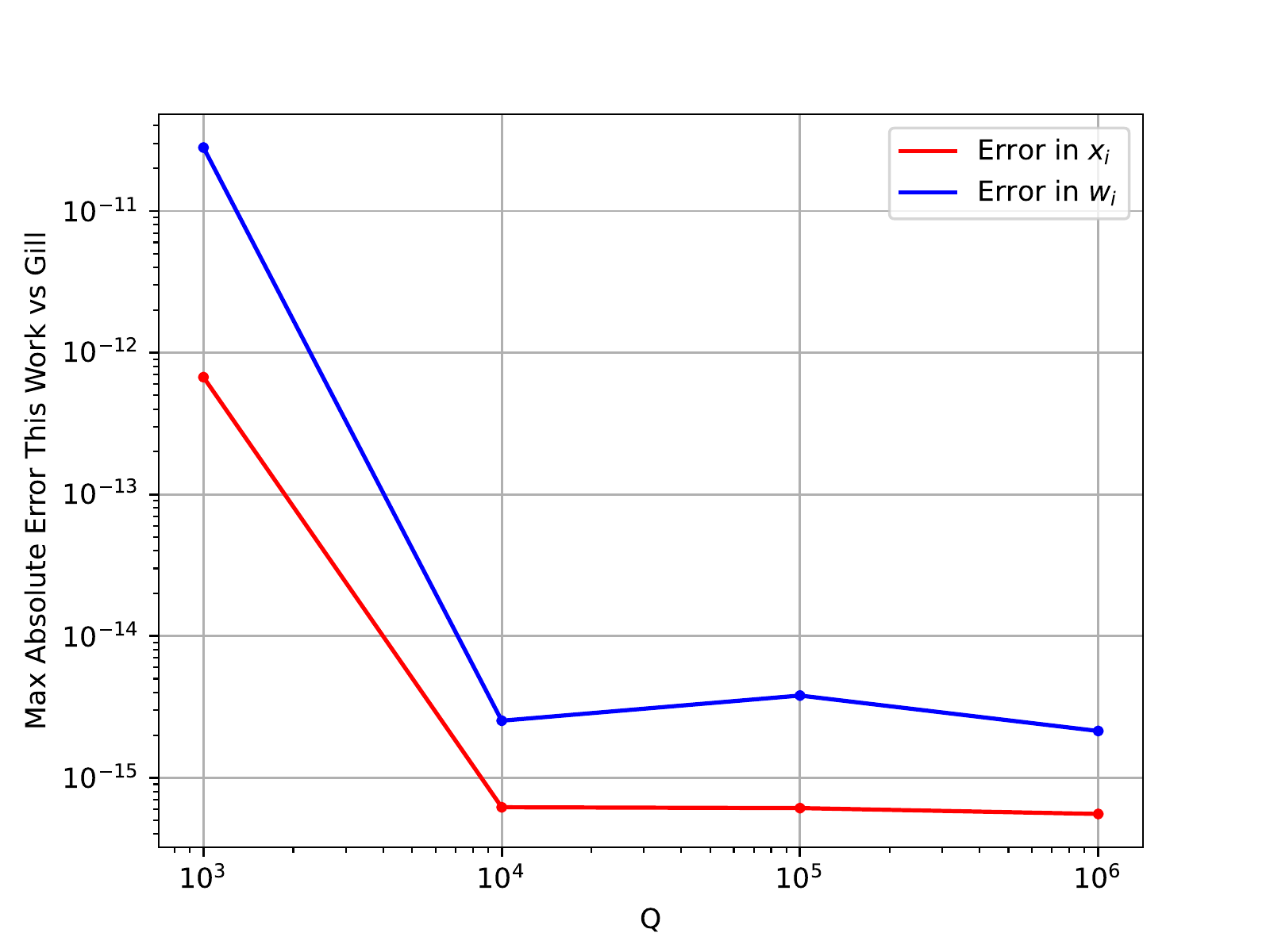}
\caption{Same as Fig. \ref{fig:delta-2}, but for $N=45$.}
\label{fig:delta-45}
\end{center}
\end{figure}

Figure \ref{fig:delta-2} shows the convergence characteristics of the $N=2$
MultiExp grid vs. the reference grid from the Gill and Chien table. This is
representative of grid convergence in the small $N$ limit. Figure
\ref{fig:delta-45} shows the same for the $N=45$ grid. The is representative of
grid convergence in the medium $N$ limit.

Both figures tell a very similar story. Regular seemingly $\mathcal{O} (Q^{-4})$
convergence occurs for both roots and weights, with the weights having a
slightly higher error prefactor. By $Q = 10^{5}$ maximum absolute
deviations of $< 10^{-15}$ in roots and $< 10^{-14}$ in weights are achieved for
these (and all other correct) grids. 

\subsection{Agreement with Gill's Grids}

Table \ref{tab:gill} shows the agreement characteristics with the tabulated Gill
and Chien MultiExp grids, for our highest-quality grids obtained with $Q =
10^{5}$. We define agreement as the maximum absolute deviation for the roots and
weights of each grid, separately. Overall, remarkable agreement is found, with
agreement of $\sim 10^{-15}$ in the roots in all cases and $\sim 10^{-14}$ in the
weights in most cases.

\begin{table}[h!]
\begin{center}
\caption{Maximum absolute deviations of MultiExp quadrature grids obtained in
this work with $Q = 10^{5}$ vs. known MultiExp quadrature grids of Gill
and Chien.}
\label{tab:gill}
\begin{tabular}{lrrrr}
\hline \hline
$N$ & 
$|\vec x - \vec x^{\mathrm{Gill}}|_{\infty}$ &
$|\vec w - \vec w^{\mathrm{Gill}}|_{\infty}$ &
Comment \\
\hline
\texttt{1 } & \texttt{  2.8E-17} & \texttt{  4.4E-16} & \\
\texttt{2 } & \texttt{  5.0E-16} & \texttt{  8.9E-16} & \\
\texttt{3 } & \texttt{  3.3E-16} & \texttt{  1.1E-15} & \\
\texttt{4 } & \texttt{  3.9E-16} & \texttt{  5.6E-16} & \\
\texttt{5 } & \texttt{  5.6E-16} & \texttt{  7.8E-16} & \\
\texttt{6 } & \texttt{  7.8E-16} & \texttt{  2.2E-15} & \\
\texttt{7 } & \texttt{  5.0E-16} & \texttt{  6.7E-16} & \\
\texttt{8 } & \texttt{  5.6E-16} & \texttt{  5.8E-16} & \\
\texttt{9 } & \texttt{  5.6E-16} & \texttt{  2.1E-15} & \\
\texttt{10} & \texttt{  4.4E-16} & \texttt{  2.2E-15} & \\
\texttt{11} & \texttt{  5.6E-16} & \texttt{  6.7E-16} & \\
\texttt{12} & \texttt{  4.3E-16} & \texttt{  2.2E-15} & \\
\texttt{13} & \texttt{  5.2E-16} & \texttt{  1.7E-15} & \\
\texttt{14} & \texttt{  6.7E-16} & \texttt{  1.2E-15} & \\
\texttt{15} & \texttt{  6.7E-16} & \texttt{  1.6E-15} & \\
\texttt{16} & \texttt{  7.2E-16} & \texttt{  1.0E-15} & \\
\texttt{17} & \texttt{  5.6E-16} & \texttt{  2.4E-15} & \\
\texttt{18} & \texttt{  6.7E-16} & \texttt{  3.4E-15} & \\
\texttt{19} & \texttt{  6.7E-16} & \texttt{  1.8E-15} & \\
\texttt{20} & \texttt{  4.7E-16} & \texttt{  1.2E-14} & \\
\texttt{21} & \texttt{  1.7E-15} & \texttt{  2.8E-15} & \\
\texttt{22} & \texttt{  4.4E-16} & \texttt{  1.1E-15} & \\
\texttt{23} & \texttt{  5.6E-16} & \texttt{  6.3E-15} & \\
\texttt{24} & \texttt{  7.8E-16} & \texttt{  4.5E-15} & \\
\texttt{25} & \texttt{  4.4E-16} & \texttt{  2.8E-15} & \\
\texttt{26} & \texttt{  5.6E-16} & \texttt{  1.9E-15} & \\
\texttt{30} & \texttt{  5.6E-16} & \texttt{  3.6E-12} & Marginal \\
\texttt{35} & \texttt{  6.7E-16} & \texttt{  3.6E-15} & \\
\texttt{40} & \texttt{  5.6E-16} & \texttt{  5.9E-15} & \\
\texttt{45} & \texttt{  6.1E-16} & \texttt{  3.8E-15} & \\
\texttt{50} & \texttt{  6.7E-16} & \texttt{  3.7E-07} & Significant \\
\hline \hline
\end{tabular}
\end{center}
\end{table}

There are two nontrivial discrepancies between our grids and those of Gill and
Chien: $N = 30$ (marginal - $4 \times 10^{-12}$) and the largest $N = 50$
(significant - $4 \times 10^{-7}$). Both occur in the weights only and both
affect almost all of the weights to the same magnitude (e.g., it is not a single
typo in a single weight). We can also rule out accidental issues with the
specific value of $Q$ used in our procedure - for these cases, the disagreement
is systematic across multiple values of $Q$.

\begin{figure*}[h!]
\begin{center}
\includegraphics[width=3.4in]{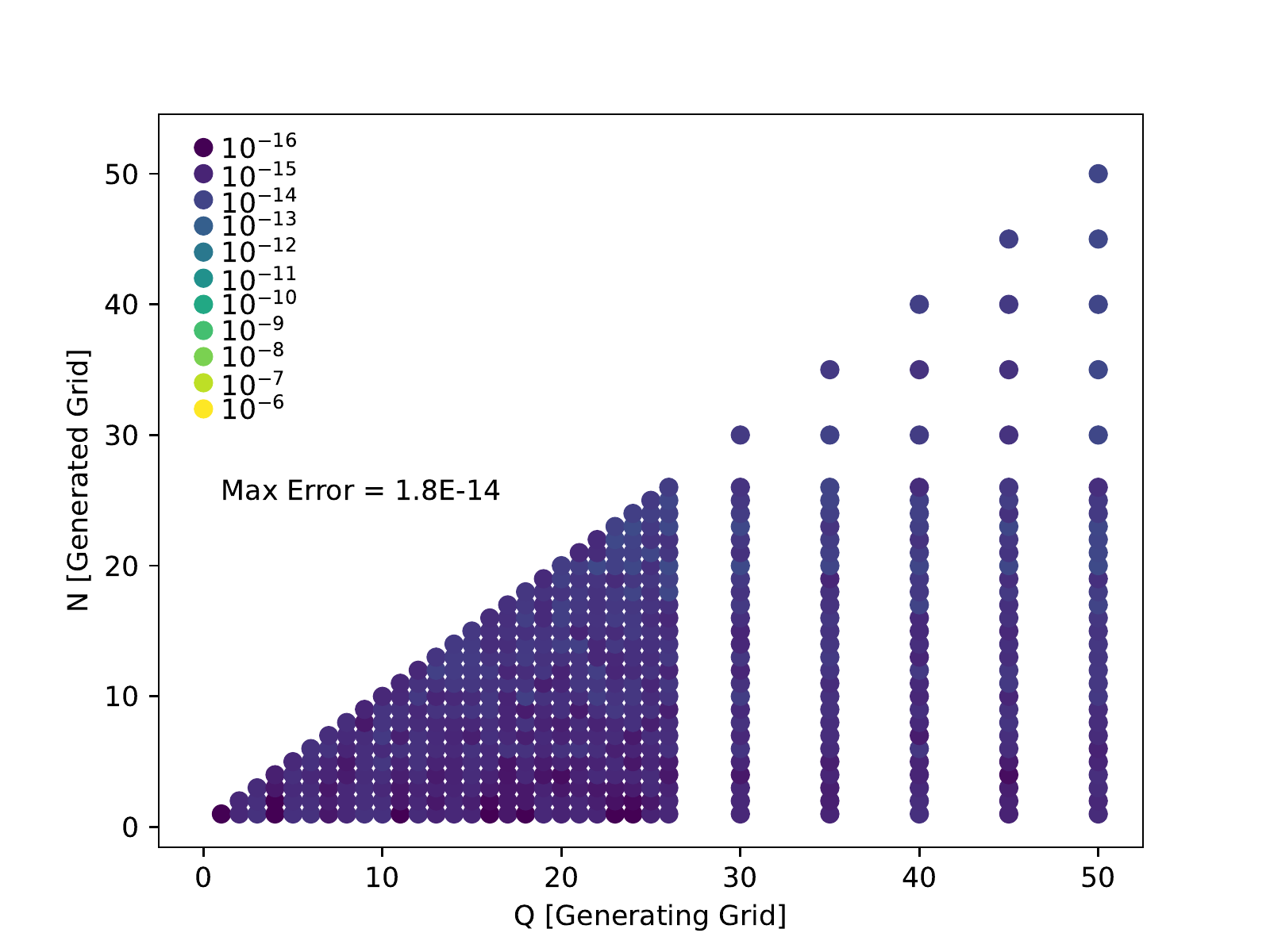}
\includegraphics[width=3.4in]{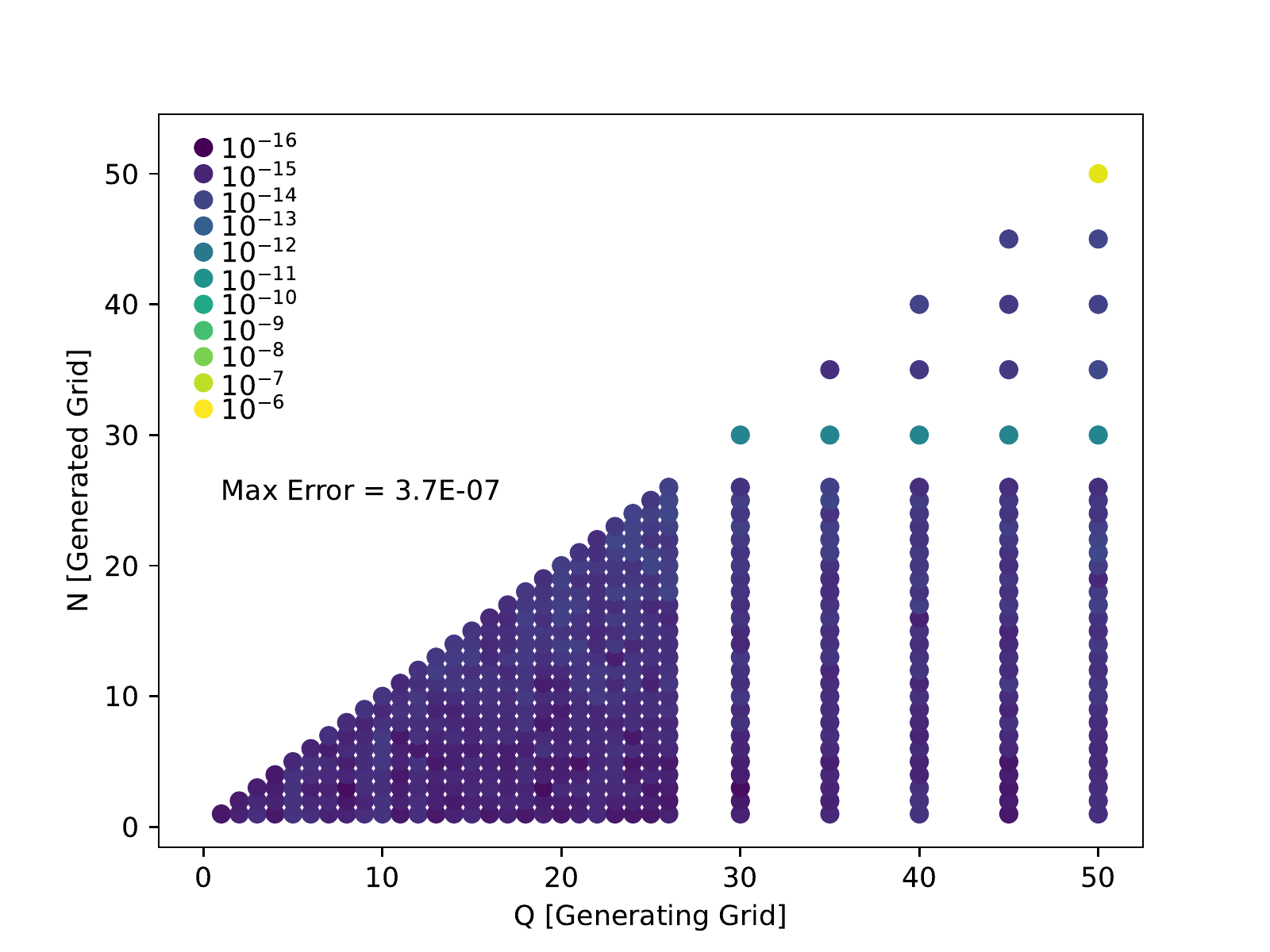}
\includegraphics[width=3.4in]{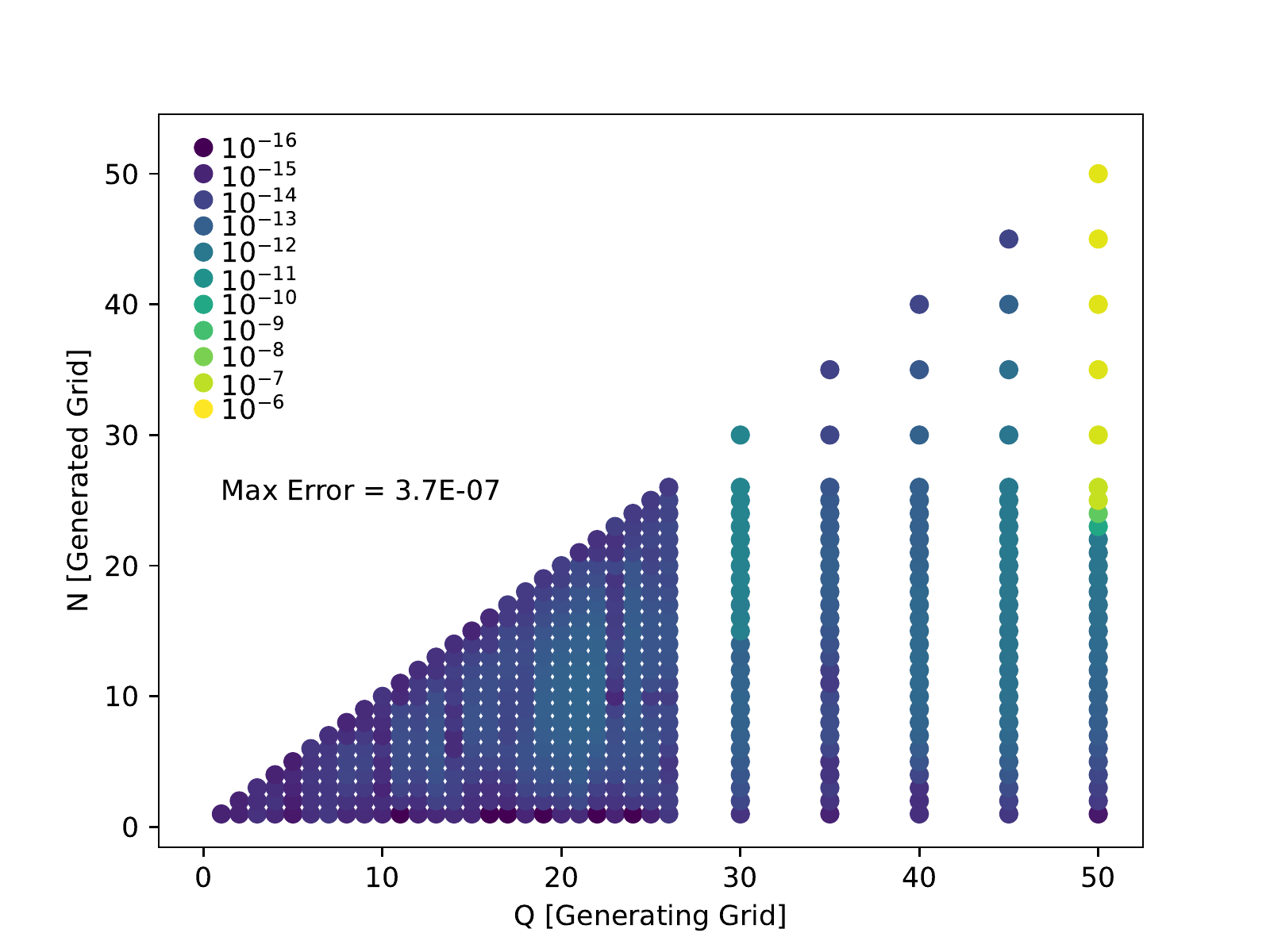}
\includegraphics[width=3.4in]{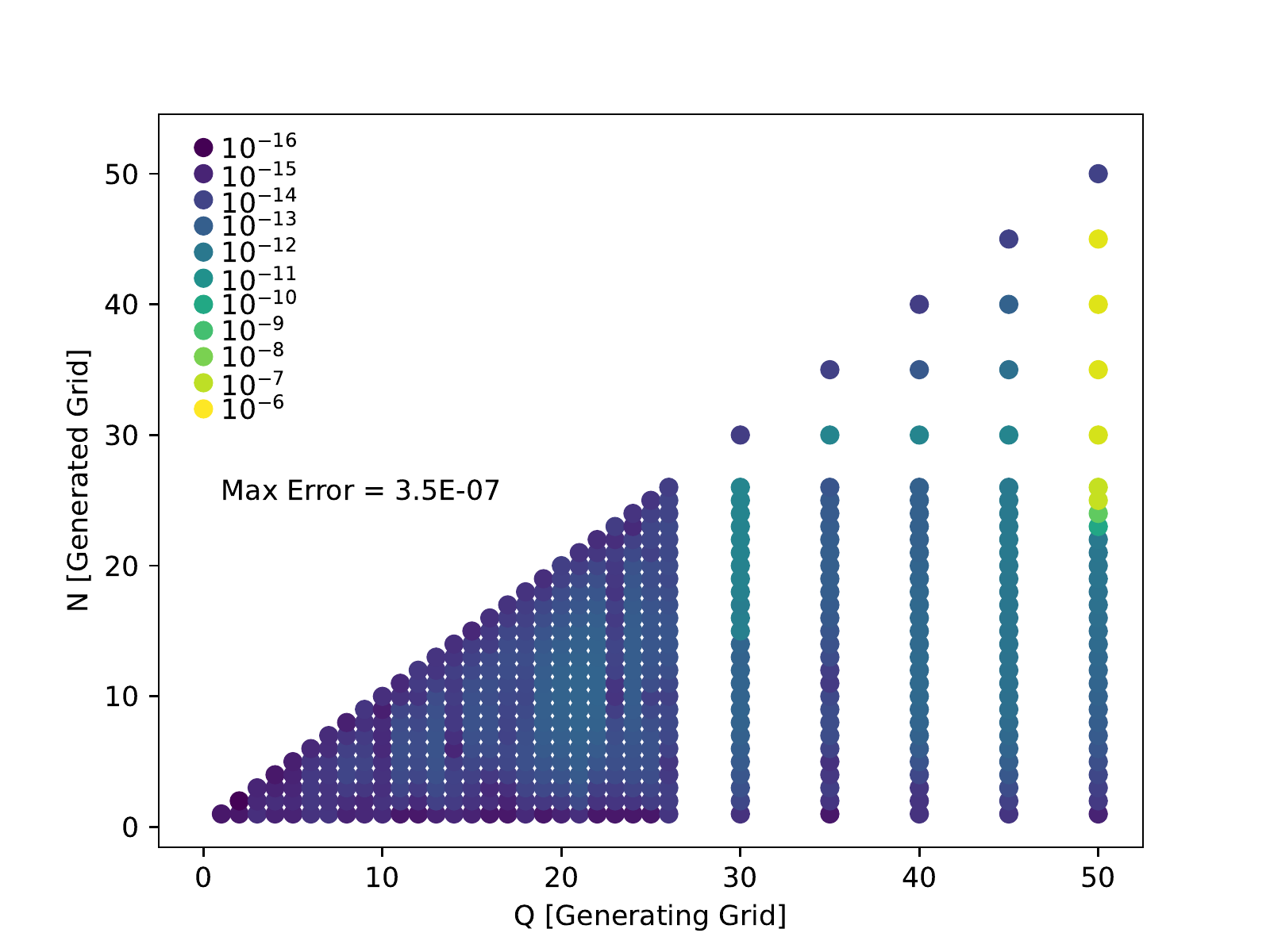}
\caption{
Maximum absolute deviation in weights generated by boosting one MultiExp
quadrature rule of size $N$ from another rule of size $Q \geq N$. 
Top row: quadrature rules generated from this work with $Q = 10^{5}$.
Bottom row: quadrature rules generated from the tables of Gill and Chien.
Left: newly generated quadrature rules compared to this work.
Right: newly generated quadrature rules compared to Gill and Chien.}
\label{fig:diff}
\end{center}
\end{figure*}

To probe deeper into these discrepancies, we use the Boley-Golub + Golub-Welsch
procedure to ``boost'' the lower MultiExp quadrature rules from the higher ones.
As a reminder, the higher MultiExp quadrature rules formally cover the lower
ones, so this procedure should be lossless. In Fig. \ref{fig:diff}, we focus on
the mean absolute deviations in the weights arising from this procedure as a
metric of accuracy - a similar story occurs in the roots. 

After a bit of analysis, the story is fairly simple. Looking at the bottom row:
The Gill and Chien $Q=50$ grid is not able to quantitatively reproduce the lower
Gill and Chien grids for $N \in \{ 25, 26, 30, 35, 40, 45 \}$. A similar, 
smaller but noticeable issue also occurs for $Q = 30$. Note that the moderate
errors (moderate blue colors) throughout the bottom row are more minor issues
that we hypothesize stem from the truncation of the smaller weights to 15
absolute decimal digits in the Gill and Chien tables, which loses a touch of
relative precision. Looking at the top row, the grids developed in this work are
all quantitatively self consistent, and can quantitatively reproduce all Gill
and Chien grids except for the $N = 50$ and to a lesser extent the $N = 30$
grids that may have problems.

\subsection{Finished MultiExp Quadrature Grids}

\begin{figure*}[h!]
\begin{center}
\includegraphics[width=3.4in]{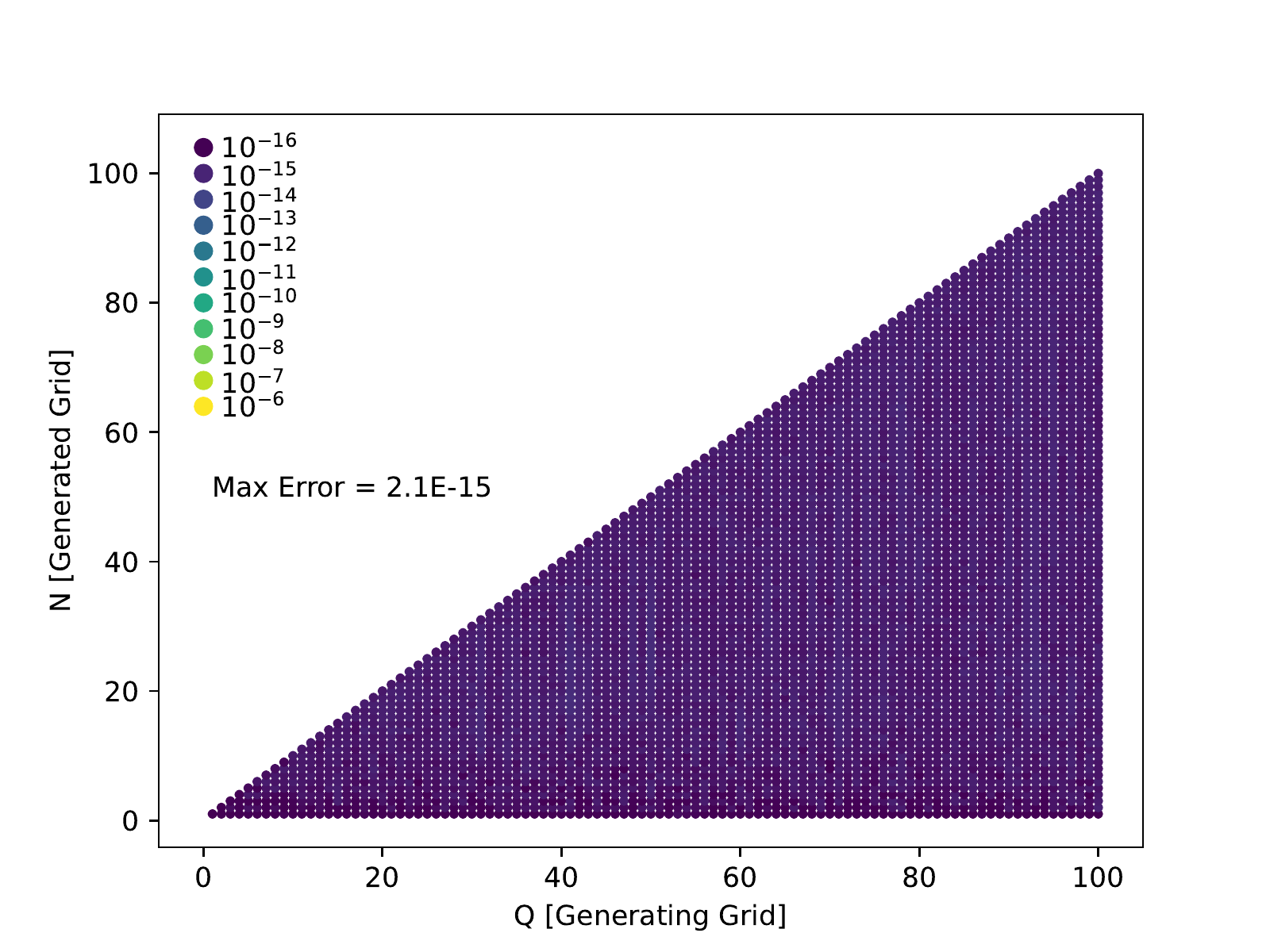}
\includegraphics[width=3.4in]{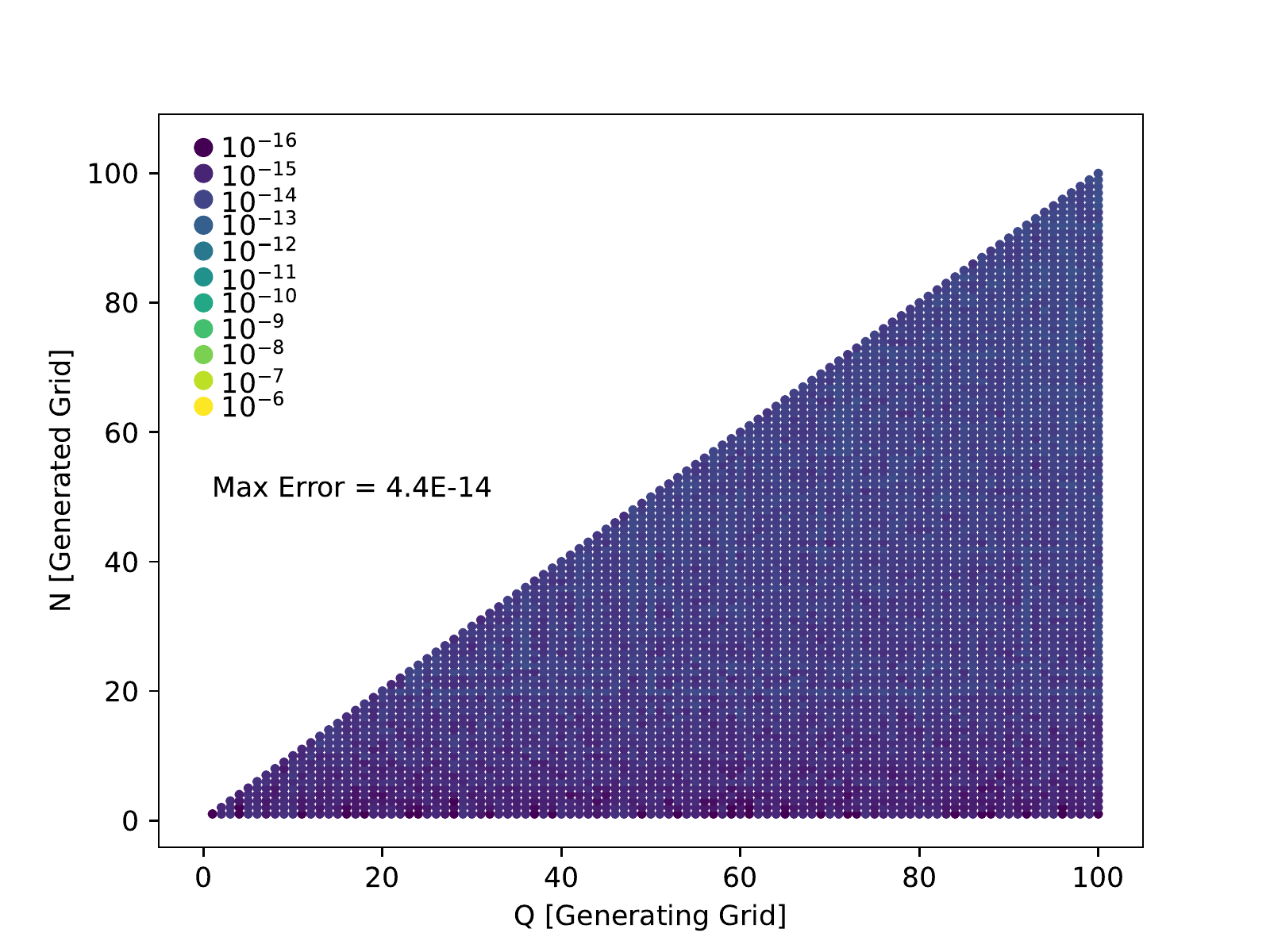}
\caption{As in Fig. \ref{fig:diff} but comparing higher MultiExp grids to
all lower MultiExp grids for the highest quality $Q = 10^{5}$ grids developed in
this work. Comparisons up to $N = 100$ are performed. Left panel: maximum
absolute deviations in roots. Right panel: maximum absolute deviations in
weights.}
\label{fig:finished}
\end{center}
\end{figure*}

The highest quality MultiExp quadrature grids computed in this work are obtained
for $1 \leq N \leq 1000$ using $Q = 10^{5}$. Fig. \ref{fig:finished} performs a
similar analysis as in the previous section, but comparing larger finished
MultiExp quadrature grids to smaller ones, up to $N=100$. Overall the story is
reassuringly boring: all lower grids are quantitatively reproduced from all
higher grids, with very slightly higher errors occurring for weights as compared
to roots.

Note that we also computed the grids up to $N = 1000$ at $Q = 10^{6}$, and find
maximum absolute deviation of $6\times 10^{-16}$ in the roots and
$3\times10^{-15}$ in the weights for all 500,500 unique grid points in this set.
We also computed the grids up to $N = 100$ at $Q = 2\times 10^{8}$ with the
slower-converging $\gamma = 1$, and find vi maximum absolute deviation of
$6\times 10^{-16}$ in the roots and $3\times10^{-15}$ in the weights for all
5,050 unique grid points in this set. All three grids are indistinguishable at
very near the double precision machine epsilon, so we have elected to use the
$Q = 10^{5}$ grid with $\gamma = 2$ as our ``highest-quality'' grid, as it is
the smallest grid that we are confident has converged, and is therefore less
susceptible to numerical noise in long-running sums and/or underflows in the
intermediates.

A visual representation of the finished MultiExp quadrature grids is shown in
Figure \ref{fig:multiexp}. As with all Gaussian quadratures over smooth
measures, the MultiExp quadrature grid exhibits a beautiful root-repulsion
property, which manifests as apparent Moir\'e patterns in the plot.

The Boley-Golub tridiagonal matrix $T_{(N=1000)}$ and all of the associated
MultiExp quadrature rules are provided in double precision \texttt{.npz} format
for the highest-quality $Q = 10^{5}$ grids obtained in this work. These can be
obtained at:
\href{https://github.com/robparrishqc/multiexp}{https://github.com/robparrishqc/multiexp}.
For the avoidance of all doubt, the complete 16-decimal-digit floating point
representation of $T_{(N=100)}$ (which can reproduce all of our grids through
Golub-Welsch for up to $N = 100$) and the MultiExp quadrature grid for $N = 100$
(for reference when others implement their own Golub-Welsch scripts) is provided
in Tables \ref{tab:N100-1} and \ref{tab:N100-2} below. 

It is extremely difficult to confidently bound all sources of errors that may be
occurring in the MultiExp grids obtained below. However, very tight agreement
with many of the Gill and Chien grids up to $N = 50$, highly regular convergence
behavior, a simple and seemingly stable algorithm, and checks of internal
Boley-Golub boosting all indicate that extremely accurate grids near the double
precision limit have been obtained. Conservatively, we estimate that these grids
are good to a maximum absolute deviation of $10^{-14}$ in the roots and
$10^{-13}$ in the weights vs. the ground truth. Note that better than one full
order of magnitude beyond this estimate was obtained for up to $N=45$ in checks
against the sensibly-behaved Gill and Chien grids. Overall, this should be well
beyond the accuracy required for density functional theory grid applications.
For applications demanding higher accuracy, testing against a massive systemic
quadrature like Gauss-Laguerre or remapped Gauss-Legendre should be performed.

\subsection{Integration Filter Functions}

To close this section, we will take a moment to explore an experimental topic:
integration ``filter functions.'' Let us consider the integration of an
arbitrary normalized Slater function with exponent $\alpha$,
\[
I (\alpha)
\equiv
\int_{0}^{\infty}
\mathrm{d} r \
r^{2}
N
\exp (-\alpha r)
=
1
\]
\begin{equation}
\approx
\sum_{i=1}^{N}
w_{i}
N 
\exp(-\alpha r_{i})
:
\
N
\equiv
\left (
\frac{2}{\alpha^3}
\right )^{-1}
\end{equation}
We can define the Slater filter function as the integration error
$\Delta^{\mathrm{Slater}} (\alpha) \equiv I (\alpha) - 1$. Here and throughout
this section we consider the nondimensional case $R = 1$.  

\begin{figure}[h!]
\begin{center}
\includegraphics[width=3.4in]{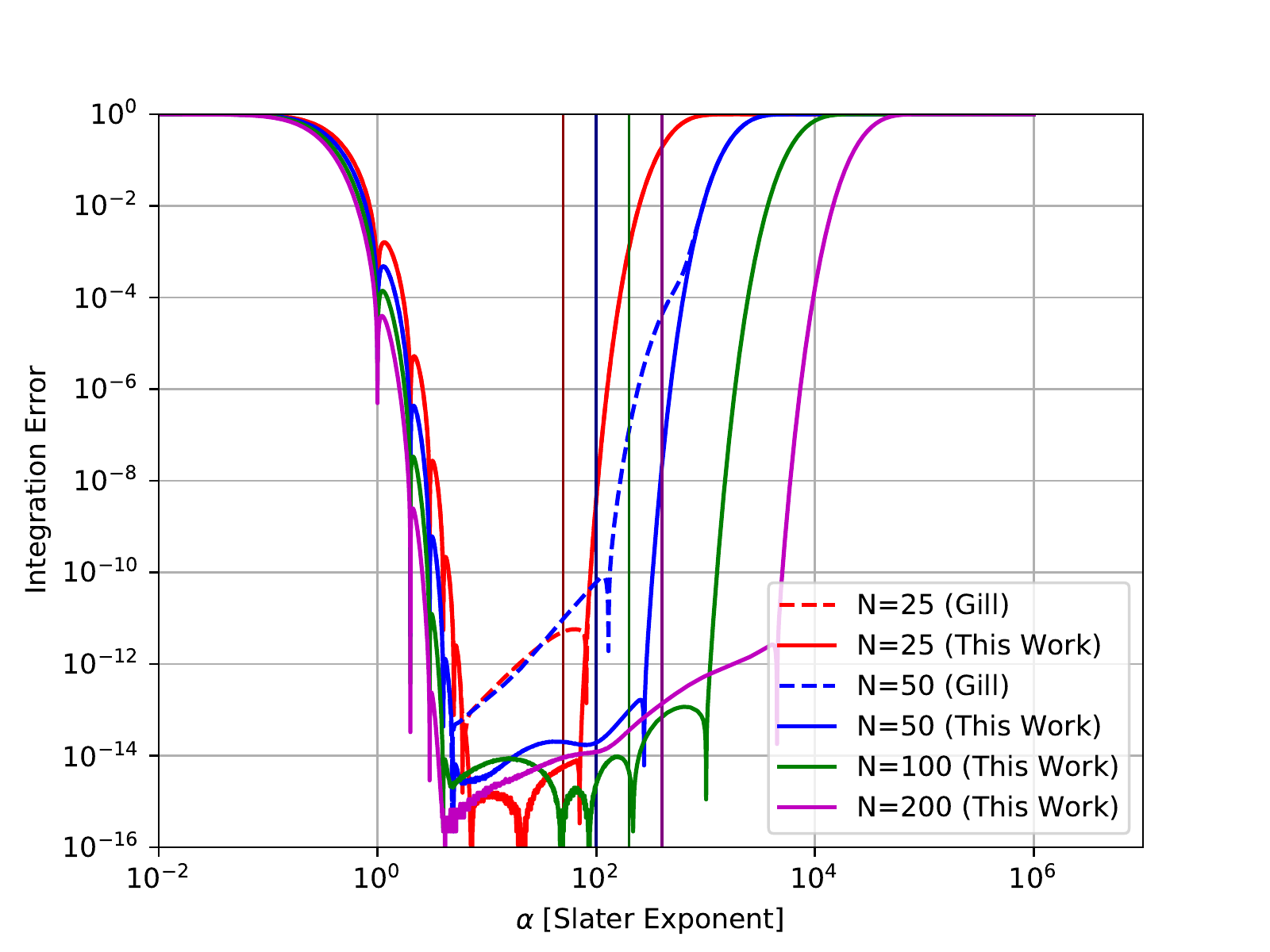}
\caption{Slater integration filter function for various MultiExp quadratures.
The plot shows the error encountered when integrating a normalized Slater
function with exponent $\alpha$, while using the dimensionless choice of $R =
1$. 
The vertical lines indicate the maximum guaranteed exactly-integrated Slater
exponent of $\alpha = 2 N$.}
\label{fig:filter-slater}
\end{center}
\end{figure}

Figure \ref{fig:filter-slater} shows the Slater filter function for several
representative $N$ using the MultiExp grids built in this work and also those of
Gill and Chien where the latter are available. Looking at the figure deeply, we
realize that there seems to be a bit of serendipity going on here with these
MultiExp quadrature grids. We are guaranteed by construction that this
quadrature rule exactly integrates $\alpha = k \ \forall \ k \in {1, \ldots,
2N}$. This is apparent from the zero crossing at $\alpha = 1$ and the integrally
spaced zeros crossings for higher $\alpha$.  However, up to this point, we have
had no guarantee or even conception of the performance for $\alpha$ points
between the integral lattice. However, the MultiExp appears to be doing
considerably more than just providing exact collocation at the integral lattice
- in fact, it seems to be exponentially suppressing the errors for intermediate
$\alpha$, leading to a large flat spot in the middle of the plot that coalesces
with errors down near the machine epsilon for a significant range of $\alpha$.
Also note that this error-suppresed region appears to promulgate to much higher
$\alpha$ than strictly conctracted by the integral lattice (highest lattice
points at $\alpha = 2N$ indicated in the darker vertical lines). 

As one minor note, most of the larger Gill and Chien quadratures exhibit the
slightly worse coalesced errors in the ``first'' region of disagreement on the
left side of the coalesced region - we attribute this to truncation errors in
the weights as reported to 15 absolute decimal digits, and find that this does not
significantly affect the filter functions for smoother integrands such as
Gaussians (see next paragraph). However, the ``second'' region of disagreement
on the right side of the coalesced region for $N=50$ is much more apparent and
is unique to the $N=50$ rule - we attribute this to the previously-mentioned
problem with this tabulated grid. Of particular note, this discrepancy poisons
the quality of the Slater integration filter in the region from $\alpha \sim
10^{+2} - 10^{+3}$, which might have significant practical ramifications. It
might well be the case that this grid is much more accurate than previously
advertised.

Another interesting case is when we consider instead a normalized Gaussian
integrand with arbitrary exponent $\beta$,
\[
J (\beta)
\equiv
\int_{0}^{\infty}
\mathrm{d} r \
r^{2}
N
\exp (-\beta r^2)
=
1
\]
\begin{equation}
\approx
\sum_{i=1}^{N}
w_{i}
N 
\exp(-\alpha r_{i}^2)
:
\
N
\equiv
4 \pi
\left (
\frac{\pi}{\beta}
\right )^{-3/2}
\end{equation}
As above, we will define the Gaussian filter function as the integration error
$\Delta^{\mathrm{Gaussian}} (\beta) \equiv J (\beta) - 1$.

\begin{figure}[h!]
\begin{center}
\includegraphics[width=3.4in]{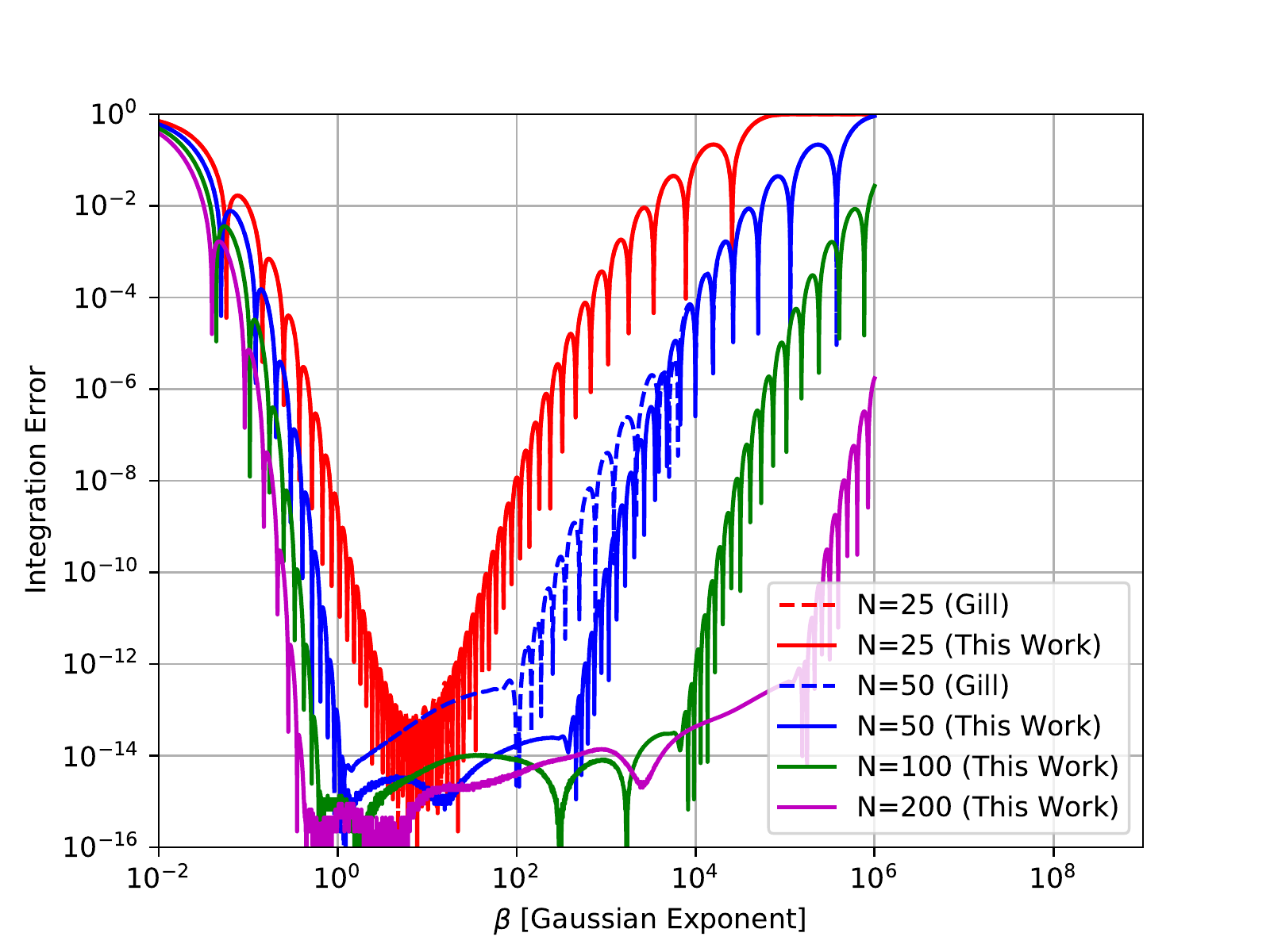}
\caption{Gaussian integration filter function for various MultiExp quadratures.
The plot shows the error encountered when integrating a normalized Gaussian
function with exponent $\beta$, while using the dimensionless choice of $R =
1$.}
\label{fig:filter-gaussian}
\end{center}
\end{figure}

Figure \ref{fig:filter-gaussian} shows the Gaussian filter function for several
representative $N$ using the MultiExp grids built in this work and also those of
Gill and Chien where the latter are available. For this case, we no longer have
any concrete idea of where exact integration is guaranteed (e.g., the zeros are
now $N$-dependent). However, a similar behavior of coalescence near the machine
precision is observed for a large range of $\beta$, with the size of this region
increasing with $N$. As promised in the last paragraph, the minor coalescence
issue with the $N=25$ Gill grid is not an issue for smoother functions like
Gaussians, as the two red curves are qualitatively indistinguishable. However,
the $N=50$ grids have significant discrepancies which poison the region of
coalescence and creep up above $10^{-6}$ integration error. Overall, the broad
accuracy of the MultiExp grids is amazing - for $N = 100$, Gaussian exponents
from $10^{0}$ to $10^{+4}$ are essentially exactly integrated in double
precision, while Gaussian exponents from $10^{-1}$ to $10^{+5}$ are integrated
to $10^{-8}$ relative precision. 

\section{Summary and Outlook}

Overall, this story is fairly simple. We have produced the MultiExp quadrature
table for all $N \leq 1000$ to what we posit is better than 14(13) decimal digits
of the maximum absolute deviations of the roots(weights). This was done by
merging the recently-developed extremely high-order Gauss-Legendre quadrature
methods of Bogaert with the Boley-Golub and Golub-Welsch algorithms. 

The combination of a high-order known quadrature rule with
Boley-Golub + Golub-Welsch in a difficult weight function may have further
runway in electronic structure and beyond. I have often used Boley-Golub in the
past for things like Rys and Boys tables, but usually with better-behaved weight
functions that only required $Q = 200$ or so. MultiExp's weight function strikes
me as unusually difficult.

As for the MultiExp grids, it may be that they deserve greater study for medium-
and high-quality density functional theory grids. The integration filter
function analysis shows compelling performance for multiscale Slater and
Guassian functions which is serendipitously beyond what is expected from even
Gaussian quadrature theory.  The broad regions of coalesced double precision
accuracy point to a loose idea that a combination of careful selection of atomic
partition function (acting as a sort of dealiasing filter) together with
well-chosen Lebedev and MultiExp spherical grids (acting as a sort of an
analog-to-digital coverter) might be able to produce extremely accurate
integration over density-involving integrands. 

It is my hope to make the simple web respository of tabulated MultExp grids
provided at
\href{https://github.com/robparrishqc/multiexp}{https://github.com/robparrishqc/multiexp}
dynamic - I will try post updates on accuracy improvements, accuracy bounds, or
rules for larger $N$ as they are obtained. 

\textbf{Acknowledgements:} Sincere thanks to I. Bogaert for writing such a clear
and marvelous paper on large-scale Gauss-Legendre quadratures. 

\textbf{Disclaimer:} RMP owns stock/options in QC Ware Corp. 

\bibliography{refs.bib}
\bibliographystyle{apsrev4-2}

\begin{table*}[h!]
\begin{center}
\caption{Table of $N=100$ MultiExp quadrature developed with the highest-quality
$Q = 10^{5}$ used in this work. Table contains both the quadrature rule, and the
diagonal / superdiagonal of the associated tridiagonal Boley-Golub matrix. The
latter can be used with the Golub-Welsch to obtain all lower quadrature rules.
Zero-based ordering is used within the table. Part 1 of 2.}
\label{tab:N100-1}
\begin{tabular}{lrrrr}
\hline \hline
$k$ & $x_{k}$ & $w_{k}$ & $T_{k+1,k+1}$ & $T_{k, k+1}$ \\
\hline
\texttt{0 } & \texttt{  1.0085604224960984E-04} & \texttt{  2.4870306762083891E-02} & \texttt{  1.2500000000000003E-01} & \texttt{  1.4142135623730949E+00} \\
\texttt{1 } & \texttt{  6.2370830580574081E-04} & \texttt{  4.1069176975224012E-02} & \texttt{  3.8851351351351354E-01} & \texttt{  1.4632852434517696E-01} \\
\texttt{2 } & \texttt{  1.6111135319247123E-03} & \texttt{  5.0506092347931963E-02} & \texttt{  4.4960948964368064E-01} & \texttt{  2.1415281692418647E-01} \\
\texttt{3 } & \texttt{  3.0668769586458212E-03} & \texttt{  5.6613009441253949E-02} & \texttt{  4.7132321389185050E-01} & \texttt{  2.3175854175497282E-01} \\
\texttt{4 } & \texttt{  4.9920519775517218E-03} & \texttt{  6.0669039914305153E-02} & \texttt{  4.8148774770913866E-01} & \texttt{  2.3895834934512467E-01} \\
\texttt{5 } & \texttt{  7.3864040114336353E-03} & \texttt{  6.3322642015236907E-02} & \texttt{  4.8705873760139035E-01} & \texttt{  2.4260216275101654E-01} \\
\texttt{6 } & \texttt{  1.0248816678655993E-02} & \texttt{  6.4957405384297978E-02} & \texttt{  4.9044203361145278E-01} & \texttt{  2.4469971118791925E-01} \\
\texttt{7 } & \texttt{  1.3577463486490179E-02} & \texttt{  6.5822739841657679E-02} & \texttt{  4.9265083085488964E-01} & \texttt{  2.4601697895485777E-01} \\
\texttt{8 } & \texttt{  1.7369896640950108E-02} & \texttt{  6.6091611891759217E-02} & \texttt{  4.9417258087122390E-01} & \texttt{  2.4689804818183456E-01} \\
\texttt{9 } & \texttt{  2.1623099391943873E-02} & \texttt{  6.5889742995765208E-02} & \texttt{  4.9526558057420966E-01} & \texttt{  2.4751623882089294E-01} \\
\texttt{10} & \texttt{  2.6333520074391794E-02} & \texttt{  6.5311842329492720E-02} & \texttt{  4.9607714413613979E-01} & \texttt{  2.4796659902484317E-01} \\
\texttt{11} & \texttt{  3.1497096119372507E-02} & \texttt{  6.4431290584450074E-02} & \texttt{  4.9669630457035846E-01} & \texttt{  2.4830479793311860E-01} \\
\texttt{12} & \texttt{  3.7109272227740756E-02} & \texttt{  6.3306241999689830E-02} & \texttt{  4.9717945800257973E-01} & \texttt{  2.4856520018151709E-01} \\
\texttt{13} & \texttt{  4.3165015005529855E-02} & \texttt{  6.1983641827715076E-02} & \texttt{  4.9756374195292374E-01} & \texttt{  2.4876995136133967E-01} \\
\texttt{14} & \texttt{  4.9658825402078673E-02} & \texttt{  6.0501968356295642E-02} & \texttt{  4.9787442418046540E-01} & \texttt{  2.4893384402785859E-01} \\
\texttt{15} & \texttt{  5.6584749772155420E-02} & \texttt{  5.8893161721076080E-02} & \texttt{  4.9812918259335398E-01} & \texttt{  2.4906706412523832E-01} \\
\texttt{16} & \texttt{  6.3936390085429201E-02} & \texttt{  5.7184015991635750E-02} & \texttt{  4.9834068656355734E-01} & \texttt{  2.4917681101791719E-01} \\
\texttt{17} & \texttt{  7.1706913627772081E-02} & \texttt{  5.5397206433533040E-02} & \texttt{  4.9851820881381631E-01} & \texttt{  2.4926828978783769E-01} \\
\texttt{18} & \texttt{  7.9889062427192889E-02} & \texttt{  5.3552062420907348E-02} & \texttt{  4.9866866281552119E-01} & \texttt{  2.4934533915605897E-01} \\
\texttt{19} & \texttt{  8.8475162565092622E-02} & \texttt{  5.1665159057777903E-02} & \texttt{  4.9879728838755616E-01} & \texttt{  2.4941084029640354E-01} \\
\texttt{20} & \texttt{  9.7457133485575775E-02} & \texttt{  4.9750777045557783E-02} & \texttt{  4.9890811555506359E-01} & \texttt{  2.4946698975587775E-01} \\
\texttt{21} & \texttt{  1.0682649738281912E-01} & \texttt{  4.7821265134587045E-02} & \texttt{  4.9900428501941568E-01} & \texttt{  2.4951548575808452E-01} \\
\texttt{22} & \texttt{  1.1657438872363589E-01} & \texttt{  4.5887329432307521E-02} & \texttt{  4.9908827375964365E-01} & \texttt{  2.4955765793855311E-01} \\
\texttt{23} & \texttt{  1.2669156394605771E-01} & \texttt{  4.3958267028700149E-02} & \texttt{  4.9916205656282697E-01} & \texttt{  2.4959455932233066E-01} \\
\texttt{24} & \texttt{  1.3716841136288330E-01} & \texttt{  4.2042156698880902E-02} & \texttt{  4.9922722347088877E-01} & \texttt{  2.4962703259914357E-01} \\
\texttt{25} & \texttt{  1.4799496129035877E-01} & \texttt{  4.0146016141017954E-02} & \texttt{  4.9928506637923215E-01} & \texttt{  2.4965575858804051E-01} \\
\texttt{26} & \texttt{  1.5916089641552880E-01} & \texttt{  3.8275932851582095E-02} & \texttt{  4.9933664371345632E-01} & \texttt{  2.4968129215845308E-01} \\
\texttt{27} & \texttt{  1.7065556241073593E-01} & \texttt{  3.6437174033842598E-02} & \texttt{  4.9938282930579375E-01} & \texttt{  2.4970408918506648E-01} \\
\texttt{28} & \texttt{  1.8246797879981239E-01} & \texttt{  3.4634279683658842E-02} & \texttt{  4.9942434973444111E-01} & \texttt{  2.4972452700576633E-01} \\
\texttt{29} & \texttt{  1.9458685007741233E-01} & \texttt{  3.2871142066682860E-02} & \texttt{  4.9946181313698762E-01} & \texttt{  2.4974292011245156E-01} \\
\texttt{30} & \texttt{  2.0700057708045777E-01} & \texttt{  3.1151074102617683E-02} & \texttt{  4.9949573165283506E-01} & \texttt{  2.4975953230313847E-01} \\
\texttt{31} & \texttt{  2.1969726860867511E-01} & \texttt{  2.9476868641935158E-02} & \texttt{  4.9952653905546629E-01} & \texttt{  2.4977458617882725E-01} \\
\texttt{32} & \texttt{  2.3266475328956535E-01} & \texttt{  2.7850850214113904E-02} & \texttt{  4.9955460471797192E-01} & \texttt{  2.4978827062800227E-01} \\
\texttt{33} & \texttt{  2.4589059168178964E-01} & \texttt{  2.6274920512261948E-02} & \texttt{  4.9958024475825580E-01} & \texttt{  2.4980074677170286E-01} \\
\texttt{34} & \texttt{  2.5936208860982973E-01} & \texttt{  2.4750598633950422E-02} & \texttt{  4.9960373099668115E-01} & \texttt{  2.4981215272064283E-01} \\
\texttt{35} & \texttt{  2.7306630572181650E-01} & \texttt{  2.3279056905569379E-02} & \texttt{  4.9962529820354401E-01} & \texttt{  2.4982260740809673E-01} \\
\texttt{36} & \texttt{  2.8699007426160728E-01} & \texttt{  2.1861152965099964E-02} & \texttt{  4.9964514999965765E-01} & \texttt{  2.4983221369819270E-01} \\
\texttt{37} & \texttt{  3.0112000804549349E-01} & \texttt{  2.0497458656735038E-02} & \texttt{  4.9966346368876136E-01} & \texttt{  2.4984106092202224E-01} \\
\texttt{38} & \texttt{  3.1544251663330630E-01} & \texttt{  1.9188286193387662E-02} & \texttt{  4.9968039423719862E-01} & \texttt{  2.4984922695883946E-01} \\
\texttt{39} & \texttt{  3.2994381868316508E-01} & \texttt{  1.7933711964510341E-02} & \texttt{  4.9969607756862916E-01} & \texttt{  2.4985677995326103E-01} \\
\texttt{40} & \texttt{  3.4460995547864703E-01} & \texttt{  1.6733598302894743E-02} & \texttt{  4.9971063330530058E-01} & \texttt{  2.4986377973943913E-01} \\
\texttt{41} & \texttt{  3.5942680461674748E-01} & \texttt{  1.5587613472126192E-02} & \texttt{  4.9972416705964079E-01} & \texttt{  2.4987027902798681E-01} \\
\texttt{42} & \texttt{  3.7438009384464777E-01} & \texttt{  1.4495250093739014E-02} & \texttt{  4.9973677235856101E-01} & \texttt{  2.4987632439976462E-01} \\
\texttt{43} & \texttt{  3.8945541503298392E-01} & \texttt{  1.3455842198038762E-02} & \texttt{  4.9974853226624999E-01} & \texttt{  2.4988195714162623E-01} \\
\texttt{44} & \texttt{  4.0463823827303552E-01} & \texttt{  1.2468581053568668E-02} & \texttt{  4.9975952075826957E-01} & \texttt{  2.4988721395220503E-01} \\
\texttt{45} & \texttt{  4.1991392608500494E-01} & \texttt{  1.1532529906098689E-02} & \texttt{  4.9976980388958997E-01} & \texttt{  2.4989212754032905E-01} \\
\texttt{46} & \texttt{  4.3526774772434540E-01} & \texttt{  1.0646637737994935E-02} & \texttt{  4.9977944079113540E-01} & \texttt{  2.4989672713433625E-01} \\
\texttt{47} & \texttt{  4.5068489357290376E-01} & \texttt{  9.8097521420505551E-03} & \texttt{  4.9978848452303570E-01} & \texttt{  2.4990103891712992E-01} \\
\texttt{48} & \texttt{  4.6615048960148436E-01} & \texttt{  9.0206313898332259E-03} & \texttt{  4.9979698280765433E-01} & \texttt{  2.4990508639909725E-01} \\
\texttt{49} & \texttt{  4.8164961189030181E-01} & \texttt{  8.2779557627929018E-03} & \texttt{  4.9980497866137641E-01} & \texttt{  2.4990889073882697E-01} \\
\hline \hline
\end{tabular}
\end{center}
\end{table*}

\begin{table*}[h!]
\begin{center}
\caption{As in Tab. \ref{tab:N100-1}. Part 2 of 2.}
\label{tab:N100-2}
\begin{tabular}{lrrrr}
\hline \hline
$k$ & $x_{k}$ & $w_{k}$ & $T_{k+1,k+1}$ & $T_{k, k+1}$ \\
\hline
\texttt{50} & \texttt{  4.9716730119366948E-01} & \texttt{  7.5803382044358865E-03} & \texttt{  4.9981251094082646E-01} & \texttt{  2.4991247101981312E-01} \\
\texttt{51} & \texttt{  5.1268857753518005E-01} & \texttt{  6.9263343434874313E-03} & \texttt{  4.9981961481651499E-01} & \texttt{  2.4991584448991430E-01} \\
\texttt{52} & \texttt{  5.2819845481955596E-01} & \texttt{  6.3144519308655101E-03} & \texttt{  4.9982632218472550E-01} & \texttt{  2.4991902676918848E-01} \\
\texttt{53} & \texttt{  5.4368195544728815E-01} & \texttt{  5.7431597272897842E-03} & \texttt{  4.9983266202668369E-01} & \texttt{  2.4992203203078781E-01} \\
\texttt{54} & \texttt{  5.5912412491814933E-01} & \texttt{  5.2108958732641373E-03} & \texttt{  4.9983866072257294E-01} & \texttt{  2.4992487315883147E-01} \\
\texttt{55} & \texttt{  5.7451004640964154E-01} & \texttt{  4.7160757688626577E-03} & \texttt{  4.9984434232678232E-01} & \texttt{  2.4992756188654330E-01} \\
\texttt{56} & \texttt{  5.8982485531643769E-01} & \texttt{  4.2570994870939436E-03} & \texttt{  4.9984972880975165E-01} & \texttt{  2.4993010891742623E-01} \\
\texttt{57} & \texttt{  6.0505375373688686E-01} & \texttt{  3.8323587415317051E-03} & \texttt{  4.9985484027097654E-01} & \texttt{  2.4993252403181032E-01} \\
\texttt{58} & \texttt{  6.2018202489268826E-01} & \texttt{  3.4402434262823332E-03} & \texttt{  4.9985969512703088E-01} & \texttt{  2.4993481618075994E-01} \\
\texttt{59} & \texttt{  6.3519504746787792E-01} & \texttt{  3.0791477441502479E-03} & \texttt{  4.9986431027790873E-01} & \texttt{  2.4993699356902291E-01} \\
\texttt{60} & \texttt{  6.5007830985333659E-01} & \texttt{  2.7474759370127787E-03} & \texttt{  4.9986870125448857E-01} & \texttt{  2.4993906372846089E-01} \\
\texttt{61} & \texttt{  6.6481742428310520E-01} & \texttt{  2.4436476308510751E-03} & \texttt{  4.9987288234953681E-01} & \texttt{  2.4994103358318787E-01} \\
\texttt{62} & \texttt{  6.7939814084888117E-01} & \texttt{  2.1661028065921165E-03} & \texttt{  4.9987686673431220E-01} & \texttt{  2.4994290950746986E-01} \\
\texttt{63} & \texttt{  6.9380636137917984E-01} & \texttt{  1.9133064068387808E-03} & \texttt{  4.9988066656255520E-01} & \texttt{  2.4994469737729122E-01} \\
\texttt{64} & \texttt{  7.0802815316976475E-01} & \texttt{  1.6837525876843753E-03} & \texttt{  4.9988429306339882E-01} & \texttt{  2.4994640261636614E-01} \\
\texttt{65} & \texttt{  7.2204976255208009E-01} & \texttt{  1.4759686240913507E-03} & \texttt{  4.9988775662452656E-01} & \texttt{  2.4994803023727005E-01} \\
\texttt{66} & \texttt{  7.3585762828657719E-01} & \texttt{  1.2885184767437005E-03} & \texttt{  4.9989106686673296E-01} & \texttt{  2.4994958487827185E-01} \\
\texttt{67} & \texttt{  7.4943839476798102E-01} & \texttt{  1.1200060278347796E-03} & \texttt{  4.9989423271089062E-01} & \texttt{  2.4995107083637444E-01} \\
\texttt{68} & \texttt{  7.6277892502972255E-01} & \texttt{  9.6907799291465755E-04} & \texttt{  4.9989726243818494E-01} & \texttt{  2.4995249209700282E-01} \\
\texttt{69} & \texttt{  7.7586631353495361E-01} & \texttt{  8.3442651567650831E-04} & \texttt{  4.9990016374439278E-01} & \texttt{  2.4995385236072101E-01} \\
\texttt{70} & \texttt{  7.8868789874175871E-01} & \texttt{  7.1479145239608001E-04} & \texttt{  4.9990294378885741E-01} & \texttt{  2.4995515506731661E-01} \\
\texttt{71} & \texttt{  8.0123127543040029E-01} & \texttt{  6.0896235264631710E-04} & \texttt{  4.9990560923875160E-01} & \texttt{  2.4995640341754122E-01} \\
\texttt{72} & \texttt{  8.1348430678065564E-01} & \texttt{  5.1578014287398909E-04} & \texttt{  4.9990816630913348E-01} & \texttt{  2.4995760039276643E-01} \\
\texttt{73} & \texttt{  8.2543513618755138E-01} & \texttt{  4.3413851944353770E-04} & \texttt{  4.9991062079925319E-01} & \texttt{  2.4995874877277999E-01} \\
\texttt{74} & \texttt{  8.3707219880405026E-01} & \texttt{  3.6298505781634868E-04} & \texttt{  4.9991297812550012E-01} & \texttt{  2.4995985115192002E-01} \\
\texttt{75} & \texttt{  8.4838423279951236E-01} & \texttt{  3.0132204463237578E-04} & \texttt{  4.9991524335134291E-01} & \texttt{  2.4996090995372314E-01} \\
\texttt{76} & \texttt{  8.5936029032304118E-01} & \texttt{  2.4820703959349040E-04} & \texttt{  4.9991742121457577E-01} & \texttt{  2.4996192744423995E-01} \\
\texttt{77} & \texttt{  8.6998974816110197E-01} & \texttt{  2.0275317420627710E-04} & \texttt{  4.9991951615213975E-01} & \texttt{  2.4996290574415617E-01} \\
\texttt{78} & \texttt{  8.8026231807912902E-01} & \texttt{  1.6412919462128724E-04} & \texttt{  4.9992153232276326E-01} & \texttt{  2.4996384683983816E-01} \\
\texttt{79} & \texttt{  8.9016805683714861E-01} & \texttt{  1.3155925600465745E-04} & \texttt{  4.9992347362764172E-01} & \texttt{  2.4996475259341247E-01} \\
\texttt{80} & \texttt{  8.9969737586980014E-01} & \texttt{  1.0432247608897793E-04} & \texttt{  4.9992534372934228E-01} & \texttt{  2.4996562475197356E-01} \\
\texttt{81} & \texttt{  9.0884105062150600E-01} & \texttt{  8.1752255773420523E-05} & \texttt{  4.9992714606911348E-01} & \texttt{  2.4996646495600530E-01} \\
\texttt{82} & \texttt{  9.1759022952794977E-01} & \texttt{  6.3235374872911318E-05} & \texttt{  4.9992888388274231E-01} & \texttt{  2.4996727474709216E-01} \\
\texttt{83} & \texttt{  9.2593644263548436E-01} & \texttt{  4.8210871351274372E-05} & \texttt{  4.9993056021510573E-01} & \texttt{  2.4996805557498722E-01} \\
\texttt{84} & \texttt{  9.3387160985062678E-01} & \texttt{  3.6168712611043006E-05} & \texttt{  4.9993217793352990E-01} & \texttt{  2.4996880880409839E-01} \\
\texttt{85} & \texttt{  9.4138804881247373E-01} & \texttt{  2.6648267649900886E-05} & \texttt{  4.9993373974007382E-01} & \texttt{  2.4996953571944552E-01} \\
\texttt{86} & \texttt{  9.4847848238176335E-01} & \texttt{  1.9236589130072258E-05} & \texttt{  4.9993524818283103E-01} & \texttt{  2.4997023753213915E-01} \\
\texttt{87} & \texttt{  9.5513604574162481E-01} & \texttt{  1.3566514639098464E-05} & \texttt{  4.9993670566634185E-01} & \texttt{  2.4997091538442229E-01} \\
\texttt{88} & \texttt{  9.6135429310711895E-01} & \texttt{  9.3145966474992437E-06} & \texttt{  4.9993811446118974E-01} & \texttt{  2.4997157035431636E-01} \\
\texttt{89} & \texttt{  9.6712720404426178E-01} & \texttt{  6.1988708890889789E-06} & \texttt{  4.9993947671286099E-01} & \texttt{  2.4997220345990456E-01} \\
\texttt{90} & \texttt{  9.7244918940592040E-01} & \texttt{  3.9764731017865003E-06} & \texttt{  4.9994079444992690E-01} & \texttt{  2.4997281566328680E-01} \\
\texttt{91} & \texttt{  9.7731509690548601E-01} & \texttt{  2.4411142694155630E-06} & \texttt{  4.9994206959160592E-01} & \texttt{  2.4997340787423311E-01} \\
\texttt{92} & \texttt{  9.8172021637864459E-01} & \texttt{  1.4204246971928530E-06} & \texttt{  4.9994330395476461E-01} & \texttt{  2.4997398095356077E-01} \\
\texttt{93} & \texttt{  9.8566028485295398E-01} & \texttt{  7.7317743420403144E-07} & \texttt{  4.9994449926039919E-01} & \texttt{  2.4997453571626216E-01} \\
\texttt{94} & \texttt{  9.8913149172461112E-01} & \texttt{  3.8640172437758307E-07} & \texttt{  4.9994565713964489E-01} & \texttt{  2.4997507293439988E-01} \\
\texttt{95} & \texttt{  9.9213048486257771E-01} & \texttt{  1.7239732240031524E-07} & \texttt{  4.9994677913934638E-01} & \texttt{  2.4997559333979216E-01} \\
\texttt{96} & \texttt{  9.9465438020399766E-01} & \texttt{  6.5660651994930723E-08} & \texttt{  4.9994786672723462E-01} & \texttt{  2.4997609762650316E-01} \\
\texttt{97} & \texttt{  9.9670078450658928E-01} & \texttt{  1.9733910354580023E-08} & \texttt{  4.9994892129673407E-01} & \texttt{  2.4997658645315632E-01} \\
\texttt{98} & \texttt{  9.9826787957629104E-01} & \texttt{  3.9883338154167677E-09} & \texttt{  4.9994994417142968E-01} & \texttt{  2.4997706044508378E-01} \\
\texttt{99} & \texttt{  9.9935496638135868E-01} & \texttt{  3.5293623678895923E-10} & \texttt{  4.9995093660922957E-01} & \texttt{  2.4997752019632452E-01} \\
\hline \hline
\end{tabular}
\end{center}
\end{table*}

\end{document}